\documentclass{pasa}%
\newcommand{\lsimeq}{{_<\atop^{\sim}}}
\newcommand{\gsimeq}{{_>\atop^{\sim}}}
\title[{\it SPICA} Photometric Survey]{Tracing the evolution of dust obscured star-formation and accretion back to the reionisation epoch with SPICA}
\author[C. Gruppioni et al.]{C. Gruppioni$^1$\thanks{email: carlotta.gruppioni@oabo.inaf.it}, L. Ciesla$^2$, E. Hatziminaoglou$^{3}$, F. Pozzi$^{4,1}$, G. Rodighiero$^{5}$, P. Santini$^{6}$, L. Armus$^7$, M. Baes$^8$,  J. Braine$^{9}$, V. Charmandaris$^{10}$, D.L. Clements$^{11}$, N. Christopher$^{12}$, H. Dannerbauer$^{13,14}$, A. Efstathiou$^{12}$, E. Egami$^{15}$, J.A. Fern\'andez-Ontiveros$^{13,14,16}$, F. Fontanot$^{17}$, A. Franceschini$^{5}$, E. Gonz\'alez-Alfonso$^{18}$, M. Griffin$^{19}$, H. Kaneda$^{20}$, L. Marchetti$^{21,22}$, P. Monaco$^{23,17}$, T. Nakagawa$^{24,25}$, T. Onaka$^{24}$, A. Papadopoulos$^{12}$, C. Pearson$^{26,20}$, I. P\'erez-Fournon$^{13,14}$, P. Per\'ez-Gonz\'alez$^{27}$, P. Roelfsema$^{28}$, D. Scott$^{29}$, S. Serjeant$^{21}$, L. Spinoglio$^{16}$, M. Vaccari$^{22}$, F. van der Tak$^{28}$, C. Vignali$^{4,1}$, L. Wang$^{28,30}$, T. Wada$^{25}$\\ \\ 
\affil{$^1$Istituto Nazionale di Astrofisica (INAF) - Osservatorio Astronomico di Bologna, via Gobetti 93/3, I--40129 Bologna, Italy.}
\affil{$^2$Laboratoire AIM-Paris-Saclay, CEA/DSM/Irfu Ð CNRS Ð Universit\'e Paris Diderot, CEA-Saclay, 91191 Gif-sur-Yvette, France.}
\affil{$^3$European Southern Observatory, Karl-Schwarzschild-Str. 2, D--85748 Garching, Germany.}
\affil{$^4$Dipartimento di Fisica e Astronomia, Universit\`a degli Studi di Bologna, Viale Berti Pichat 6/2, 40127 Bologna, Italy.}
\affil{$^5$Dipartimento di Fisica e Astronomia ``G. Galilei'', Universit\`a di Padova, Vicolo dell'Osservatorio 3, 35122, Italy.}
\affil{$^6$INAF - Osservatorio Astronomico di Roma, via di Frascati 33, 00078, Monte Porzio Catone, Italy.}
\affil{$^7$Spitzer Science Center, California Institute of Technology, Pasadena, CA, USA.}
\affil{$^8$Sterrenkundig Observatorium, Universiteit Gent, Krijgslaan 281 S9, 9000, Gent, Belgium.}
\affil{$^9$Observatoire de Bordeaux, Laboratoire d'Astrophysique de Bordeaux, 2 rue de l'Observatoire, BP 89, 33270 Floirac, France.}
\affil{$^{10}$Institute for Astronomy \& Astrophysics, Space Applications \& Remote Sensing, National Observatory of Athens, Palaia Penteli 15236, Athens, Greece.}
\affil{$^{11}$Blackett Lab, Imperial College, London, Prince Consort Road, London SW7 2AZ, UK.}
\affil{$^{12}$School of Sciences, European University Cyprus, Diogenes Street, Engomi, 1516 Nicosia, Cyprus.}
\affil{$^{13}$Instituto de Astrof\'isica de Canarias, C/V\'ia L\'actea, s/n, E-38205 La Laguna, Tenerife, Spain.}
\affil{$^{14}$Universidad de La Laguna, Dept. de Astrof\'isica, C/Astrof\'isico Fco. S\'anchez s/n, E--38206 La Laguna, Spain.}
\affil{$^{15}$Steward Observatory, University of Arizona, 933 N. Cherry Ave, Tucson, AZ 85721, USA.}
\affil{$^{16}$Istituto di Astrofisica e Planetologia Spaziali, INAF, Via Fosso del Cavaliere 100, 00133 Roma, Italy.}
\affil{$^{17}$INAF - Osservatorio Astronomico di Trieste, via G.B. Tiepolo 11, I-34143 Trieste, Italy.}
\affil{$^{18}$Universidad de Alcal\'a, Departamento de F\'isica y Matem\'aticas, Campus Universitario, 28871 Alcal\'a de Henares, Madrid, Spain.}
\affil{$^{19}$School of Physics and Astronomy, Cardiff University, The Parade, Cardiff CF24 3AA, UK.}
\affil{$^{20}$Graduate School of Science, Nagoya University, Furo-cho, Chikusa-ku, Nagoya 464-8602, Japan.}
\affil{$^{21}$School of Physical Sciences, The Open University, Milton Keynes, MK7 6AA, UK.}
\affil{$^{22}$Department of Physics and Astronomy, University of the Western Cape, R. Sobukwe Road, 7535 Bellville, Cape Town, South Africa.}
\affil{$^{23}$Dipartimento di Fisica - Sezione di Astronomia, Universit\'a di Trieste, Via Tiepolo 11, 34131 Trieste, Italy.}
\affil{$^{24}$Department of Astronomy, Graduate School of Science, The University of Tokyo, 7-3-1 Hongo, Bunkyo-ku, Tokyo 113-0033, Japan.}
\affil{$^{25}$Institute of Space Astronautical Science, Japan Aerospace Exploration Agency, 3-1-1 Yoshinodai, Chuo-ku, Sagamihara, Kanagawa 252-5210, Japan.}
\affil{$^{26}$RAL Space, CCLRC Rutherford Appleton Laboratory, Chilton, Didcot, Oxfordshire OX11 0QX, UK.}
\affil{$^{27}$Departamento de Astrof\'isica, Facultad de CC. F\'isicas, Universidad Complutense de Madrid, 28040 Madrid, Spain.}
\affil{$^{28}$SRON Netherlands Institute for Space Research, Landleven 12, 9747 AD, Groningen, The Netherlands.}
\affil{$^{29}$University of British Columbia, Physics \& Astronomy Dept., 6224 Agricultural Road, V6T 1Z1 Vancouver, Canada.}
\affil{$^{30}$Kapteyn Astronomical Institute, University of Groningen, Postbus 800, 9700 AV, Groningen, The Netherlands.}
}
\jid{PASA}
\doi{10.1017/pas.\the\year.xxx}
\jyear{\the\year}

\usepackage[authoryear]{natbib}
\bibpunct{(}{)}{;}{a}{}{,}
\usepackage{aas_macros}
\usepackage{hyperref} 
\hypersetup{colorlinks,citecolor=blue,linkcolor=blue,urlcolor=blue}

\begin{document}%
\begin{abstract}
Our current knowledge of star formation and accretion luminosity at high-redshift ($z$$>$3--4), as well as the possible connections between them, relies mostly on observations in the 
rest-frame ultraviolet (UV), which are strongly affected by dust obscuration.
Due to the lack of sensitivity of past and current infrared (IR) instrumentation, so far it has not been possible to get a glimpse into the early phases of the dust-obscured Universe.
Among the next generation of IR observatories, SPICA, observing in the 12--350\,$\mu$m range, will be
the only facility that can enable us to make the required leap forward in understanding the obscured star-formation rate and black-hole accretion rate densities (SFRD and BHARD, respectively) 
with respect to what {\em Spitzer} and {\em Herschel} achieved in the mid- and far-IR at $z$$<$3. In particular, SPICA will have the unique ability to trace the evolution of the obscured SFRD and BHARD
over cosmic time, from the peak of their activity back to the reionisation epoch (i.e., 3$<$$z$$\lesssim$6--7), where its predecessors had severe limitations. 
Here we discuss the potential of both deep and shallow photometric surveys performed with the SPICA mid-IR instrument (SMI), enabled by the very low level of impact of dust obscuration in a band 
centred at 34\,$\mu$m. These unique unbiased photometric surveys that SPICA will perform will be followed up by observations both with the SPICA spectrometers and with other facilities at shorter and longer wavelengths, with the aim to fully characterise the evolution of AGNs and star-forming galaxies after re-ionisation.
\end{abstract}
\maketitle
{\bf Preface}

\vspace{0.5cm}
\noindent

The following set of papers describe in detail the science goals of the future Space Infrared telescope for Cosmology and Astrophysics (SPICA). The SPICA satellite will employ a 2.5-m telescope, actively cooled to around 6\,K, and a suite of mid- to far-IR spectrometers and photometric cameras, equipped with state of the art detectors. In particular the SPICA Far Infrared Instrument (SAFARI) will be a grating spectrograph with low (R$=$300) and medium (R$\simeq$3000--11000) resolution observing modes instantaneously covering the 35--230\,$\mu$m wavelength range. The SPICA Mid-Infrared Instrument (SMI) will have three operating modes: a large field of view (12$^{\prime}$$\times$10$^{\prime}$) low-resolution 17--36\,$\mu$m spectroscopic (R$\sim$50--120) and photometric camera at 34\,$\mu$m, a medium resolution (R$\simeq$2000)  grating spectrometer covering wavelengths of 17--36\,$\mu$m and a high-resolution echelle module (R$\simeq$28000) for the 12--18\,$\mu$m domain. A large field of view (80$^{\prime \prime}$$\times$$80^{\prime \prime}$), three channel, (110\,$\mu$m, 220\,$\mu$m and 350\,$\mu$m) polarimetric camera will also be part of the instrument complement. These articles will focus on some of the major scientific questions that the SPICA mission aims to address, more details about the mission and instruments can be found in Roelfsema et al. 2017, submitted to A\&A.

\section{INTRODUCTION }
\label{sec:intro}
One of the most important themes in extragalactic astronomy over the next decades will be the exploration of the very early stages of galaxy formation, when the first light arose from the so-called 
Dark Ages, with many future facilities planning to dedicate a significant effort in pursuing this goal. 
The properties and evolution of galaxies appear to be linked to the growth of their central black holes (e.g., \citealt{kormendy13}, and references therein), 
although the mechanisms responsible for this link are not understood: how supermassive black holes (SMBHs)
formed at the centers of galaxies, and what their role was in the formation and evolution of galaxies, will still be key astrophysics questions in the coming years.
There is observational evidence, as well as a theoretical rationale to suggest an intimate relation between the evolution of galaxies and the growth of their central SMBHs. 
On the theoretical side, cosmological simulations require feedback from active galactic nuclei (AGN) in order to suppress star formation (SF) in massive galaxies 
(e.g., \citealt{dimatteo05}, \citealt{croton06}, \citealt{booth09}). 
Observational evidence in the local Universe includes the very tight relation between the black hole (BH) mass and the stellar velocity dispersion ($M_{\rm BH}$-{$\sigma$}$_{*}$; e.g., 
\citealt[and references therein]{gultekin09}) and the molecular outflows that are omni-present in local Ultra-Luminous InfraRed Galaxies (ULIRGs; e.g., \citealt{spoon13}). 
The tight $M_{\rm BH}$-{$\sigma$}$_{*}$ relation, however, breaks early on (as early as $z$$\simeq$0.3; see \citealt{woo08}), and while at high $z$ ionised-gas outflows are very common, there 
are only a handful of {\it tentative} detections of molecular outflows (e.g., \citealt{falgarone15}). Therefore, the nature of the symbiotic relation between galaxy evolution and 
accretion onto SMBHs is still matter of debate.
For this reason, understanding the formation of galaxies and their subsequent evolution will be possible only if we obtain coeval observations of the emitted radiation due to accretion of matter 
onto the first SMBHs that formed in the Universe. We therefore need to uncover and cover galaxies and active galactic nuclei (AGN) over a broad range of redshifts and luminosities, in order to 
characterise the accretion and SF histories, and to be able to separate the two contributions within the same sources.

The bulk of the SF and SMBH accretion in galaxies took place more than six billion years ago, sharply dropping towards the present epoch (e.g., \citealt{madau14}).
Most of the energy emitted by stars and accreting SMBHs is absorbed by dust, and re-emitted at longer wavelengths. 
The existence of a large number of distant sources radiating the bulk of their energy in the IR implies that the critical phases of SF and black hole (BH) accretion history took place in 
heavily obscured systems, embedded within large amounts of gas and dust (e.g., \citealt{burgarella13}; \citealt{madau14}). 
The substantial reddening affecting these dust-obscured objects makes their characterisation in the optical/UV severely biased and sometimes even impossible
(e.g., \citealt{rowanrobinson97} and \citealt{hughes98} first showed that the star-formation rate density, SFRD, was higher when measured in the far-IR and sub-mm 
than in the optical/UV). 
By observing and measuring dust re-emission in the IR, we will indirectly measure the primary radiation, looking at the main processes at work in obscured regions of galaxies.
Determining how much of the star-formation rate and black-hole accretion rate densities (SFRD and BHARD, respectively) are obscured by dust between 
$z$$\simeq$3 and $z$$\simeq$6 will have also important implications for our 
knowledge of the dust content within galaxies, as well as for the dust attenuation law evolution and, consequently, for the epoch and processes of dust formation. 

The advent of space-based infrared (IR) observatories allowed us to directly
measure the dust-obscured SF activity, independently on any extinction corrections. 
Evolutionary studies with deep {\em Spitzer} mid-IR (24-$\mu$m) surveys 
first unveiled the dust-obscured SF to $z$$\sim$2.5--3 (i.e., \citealt{lefloch05}; \citealt{perezgonzalez05}; \citealt{caputi07}); \citet{rodighiero10} -- albeit with large uncertainties at $z$$>$1.5--2
due to the difficulty in recovering the far-IR bump at longer wavelengths. 
The deepest {\em Herschel} surveys in the far-IR with PACS (observing at 100 and 160\,$\mu$m; \citealt{poglitsch10}) allowed us to trace the SFRD and BHARD evolution up to $z$$\sim$3--4 
(i.e., \citealt{gruppioni13, magnelli13, delvecchio14}), while those with the longer wavelength instrument (SPIRE, observing at 250, 350 and 500\,$\mu$m; \citealt{griffin10}), detected higher 
redshift galaxies (to $z$$\gsimeq$6; e.g., \citealt{riechers13}, \citealt{lutz14}, \citealt{rowanrobinson16}, \citealt{laporte17}), but with large identification uncertainties due to source blending.
The first far-IR luminosity function (LF) reaching $z$$\simeq$4 (\citealt{gruppioni13}) obtained with the {\em Herschel} PACS Evolutionary Probe (PEP; \citealt{lutz11}) survey data, 
suggested a rapid decline of the SFRD at $z$$>$3, following a plateau between $z$$\sim$1 and $z$$\sim$3. A decline is suggested in the SFRD at $z$$>$3, confirmed by the recent
{\it ``super-deblending''} analysis of deep {\em Herschel} images by \citet{liu17} in the GOODS-North field. The latter authors found that there is still a significant contribution from 
the dusty galaxies missed by optical to near-IR colour selections at 3$<$$z$$<$6. 
However, a recent analysis of {\em Herschel}~500-$\mu$m sources in the {\em Herschel} Multi-tiered Extragalactic Survey (HerMES; \citealt{oliver12}) by \citet{rowanrobinson16} 
has provided hints of even higher SFRD at $z$$=$3--6 (significantly higher than UV estimates), with the obscured SFRD remaining almost constant up to 
$z$$\sim$6, while the UV estimates show a decline of more than an order of magnitude towards higher redshifts.  
This result, in agreement with the previous work by \citet{dowell14}, and with the GRB measurements by \citet{kistler09}, seems to imply that the epoch of high SFRD, and hence of rapid heavy element formation, extends at least from 
$z$$\sim$6 to $z$$\sim$1. This is a significantly earlier start to the epoch of high SFRD than inferred from previous studies and could pose problems for theoretical models of galaxy formation (setting the epoch of intense activity at $z$$\sim$1--3). Moreover, this result strengthens the inconsistency between IR data and semi-analytic models (SAMs)
at high-$z$ previously found by \citet{gruppioni15}, with SAMs largely under-predicting the high SFRs seen in starburst galaxies at $z$$>$2. 

On the other hand, recent Atacama Large Millimeter Array (ALMA) observations of the Hubble Ultra Deep Field (\citealt{dunlop17}) seem to contradict these results, by finding that the SFRD peaks at $z$$\simeq$2.5, and that 
the star-forming Universe transits from primarily unobscured to primarily obscured at $z$$\simeq$4. Note that this result depends strongly on the
template SEDs adopted to compute the SFR, since ALMA samples the rest-frame SED significantly longward of the far-IR peak
(although \citealt{dunlop17} fit the template considering the {\em Herschel} limits, and estimate a maximum difference of  20-30\% due to this).
The \citet{dunlop17} result seems to be in agreement with recent ALMA observations of the Hubble Frontier Fields to extremely deep limits,  
suggesting the presence of very little dust in galaxies at $z$$>$4 (Bauer et al. in preparation), and with the result by \citet{Bourne17}. The latter authors find a transition between obscured and 
unobscured SFRDs at $z$$\sim$3, with a high redshift decline following that of UV surveys, by stacking SCUBA-2 images up to $z$$\sim$5.

In summary we have on the one hand, the SPIRE large-scale (e.g., a total of $>$800 deg$^2$ surveyed; \citealt{oliver12}, \citealt{eales10}, \citealt{valiante16}) and meaningful statistics 
(several hundreds of thousands sources detected) at high redshift that are affected by identification and redshift measurement problems, while on the other hand, 
the more precise ALMA results that are based on very small statistics and sky area coverage (e.g., the largest continuum survey of contiguous fields performed with ALMA 
covers only 4.5 arcmin$^2$ and has detected only 16 sources; \citealt{dunlop17}).
\begin{figure*}
\begin{center}
\includegraphics[width=15cm]{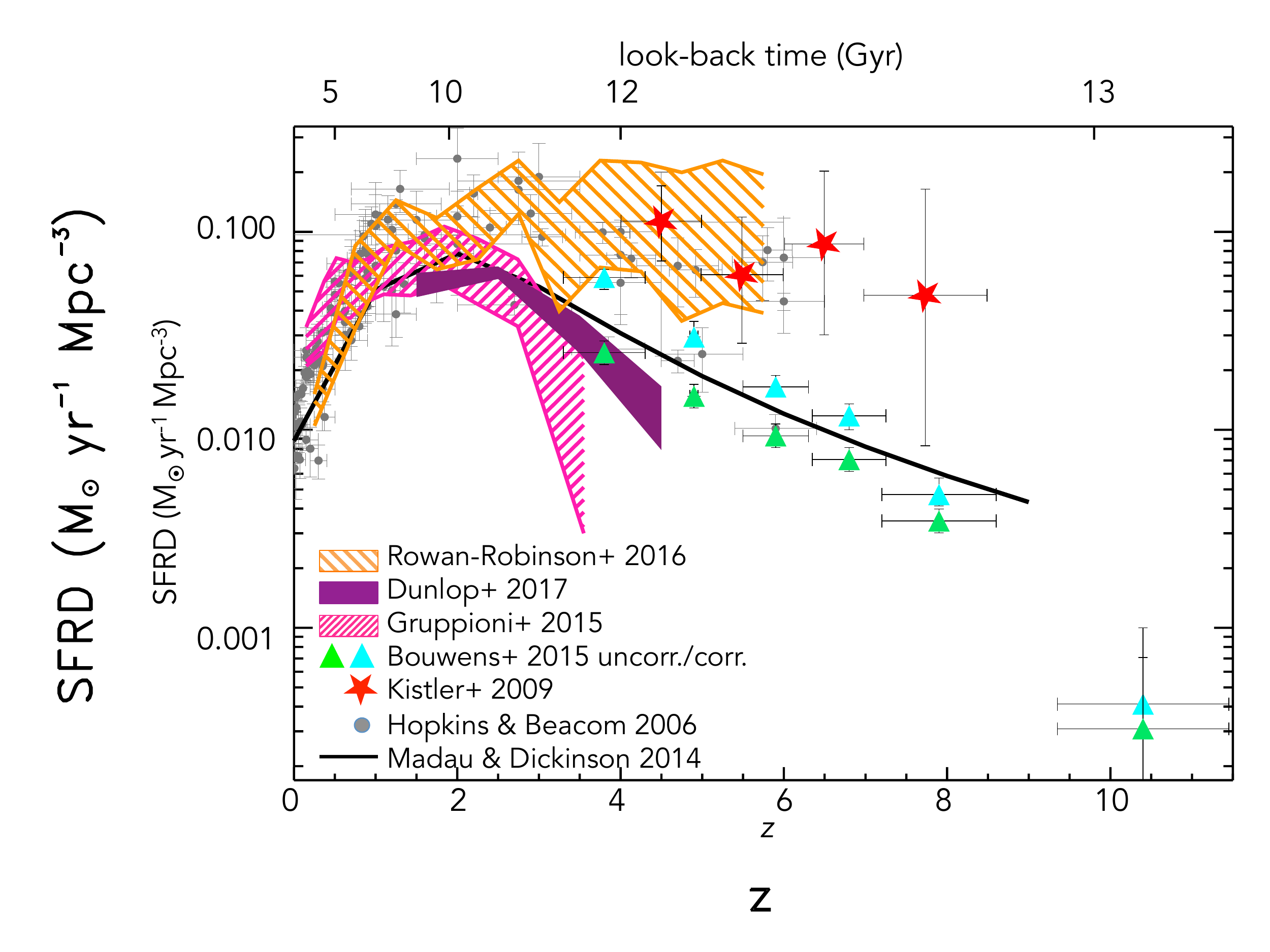}
\caption{Redshift evolution of the comoving SFRD. Different derivations of the obscured and unobscured SFRD are compared: the light grey data points show the extinction-corrected optical/UV compilation 
by \citet{hopkins06}; the pink hatched area shows the IR SFRD from \citet{gruppioni15} (obtained by integrating the {\em Herschel} LF of \citet{gruppioni13} after subtracting the AGN contribution from each source); the orange hatched area is the SFRD obtained by \citet{rowanrobinson16} by re-analysing the {\em Herschel} high-$z$ sub-mm sources; the purple area shows the uncertainty region of the \citet{dunlop17} SFRD from {\em ALMA} data; the black solid line shows the best-fit model by \citet{madau14} to the dust-corrected UV data and IR data; the cyan (green) filled triangles are the dust-corrected (uncorrected) UV data by \citet{bouwens15}; and the red stars are the measurements derived from high-$z$ GRB by \citet{kistler09}.}
\label{sfrdz}
\end{center}
\end{figure*} 
In Figure~\ref{sfrdz} we show our current knowledge (or rather lack of it) of the cosmic history of the SFRD, as derived by different surveys (most of which have been discussed above) performed at different wavelengths. The large spread in the results and estimates at $z$$>$3 is clearly evident, not only between obscured (e.g., IR/sub-mm) and unobscured (e.g., optical/UV) survey data, but also between IR/sub-mm surveys in different fields, with different instruments (e.g., \citealt{rowanrobinson16} with {\em Herschel} and \citealt{dunlop17} with ALMA).
Note that at $z$$\gtrsim$6 the current estimates differ by more than an order of magnitude, with very different implications for galaxy formation and evolution scenarios (see, e.g., \citealt{fontanot17}).
However, as evident from Figure~\ref{sfrdz}, at $z$$\gtrsim$3 there are currently no IR surveys that can 
conclusively confirm whether the obscured SFRD declines (and whether how) from $z$$\sim$1.5, as suggested by UV studies, or remains constant to 
$z$$\sim$6, as suggested from gamma-ray burst (GRBs; \citealt{kistler09}) and sub-mm measurements (e.g., \citealt{rowanrobinson16}).

Indeed we have very little direct evidence characterising dusty SF at very high redshifts, since current estimates of the SFRD and SF population 
at $z$$>$3--4 are nearly all derived by the observation of unobscured objects from rest-frame UV or optical surveys (e.g., \citealt{bouwens15}). 
\citet{bouwens09} found no relevant dust attenuation at $z$$>$5 (based on the UV continuum slope of high-$z$ Lyman Break Galaxies, LBGs), and they concluded 
that the UV luminosity function (LF) contains all the information 
needed to estimate the SFRD at these redshifts. Nonetheless, this might represent only a partial view: 
in fact, if dusty obscured objects do exist at high redshift,
they would be totally missed by the UV surveys and their contribution will remain completely unknown without far-IR surveys. 

From the AGN side, all the available estimates of the BHARD are at $z$$<$3. They include X-ray (e.g., \citealt{merloni07}), optical (e.g., \citealt{hopkins07}) and IR survey data (e.g., \citealt{delvecchio14}).
All of them are strongly dependent on the bolometric correction (which can be as high as a factor of 100 in the X-ray) and subject to large uncertainty about the contribution of heavily-obscured 
and Compton-thick (CT; $N_{\rm H}$$>$10$^{24}$ cm$^{-2}$) AGN. The latter can severely affect the optical and X-ray measurements,  
especially at high-$z$, where larger fractions of CT AGN are expected. 
Indeed, these sources are responsible for most of the power produced by accretion in the Universe 
(e.g., \citealt{gilli07}; \citealt{treister09}; \citealt{comastri15}; \citealt{harrison16}) and therefore are likely to represent a crucial phase in the joint evolution of galaxies and AGN. 
Although they are expected to make up a significant fraction of the X-ray background (e.g., \citealt{gilli07}), the fainter fraction of the CT-AGN population
escapes even the deepest X-ray surveys currently available. 
However, thanks to the reprocessed IR emission from the circum-nuclear dusty torus obscuring the optical/UV/soft X-ray, these objects are expected to be detectable in IR surveys. 

Observations in the mid-IR to very faint flux densities (around a few $\mu$Jy) are the only means to detect SF and/or AGN activity in most of these heavily obscured sources at large cosmological distances. Indeed, a significant fraction of the star-forming galaxies dominating the SFRD of the Universe at $z$$=$1--4, when most of the stellar populations in the most massive galaxies were being assembled (\citealt{fontana06, perezgonzalez08, marchesini09}), are very faint in the UV/optical (\citealt{perezgonzalez05}), and their SFRs can only be measured through mid-/far-IR observations reaching fluxes well below 1 mJy ({\em Herschel} reached the confusion limit of $\simeq$1 mJy at 100\,$\mu$m, and so was unable to go much further than $z$$\sim$3--4).
In this context, a deep photometric survey in the IR, reaching up to $z$$\sim$6--7, will be crucial for clarifying the evolution of the dust-obscured SF and accretion activity, allowing us to reveal, for the first time, the dark side of the reionisation epoch and of the newly formed galaxies and black holes.
The deepest extragalactic surveys performed by {\em Spitzer}-MIPS and {\em Herschel}-PACS (e.g., GOODS, H-GOODS and PEP) could not reach redshifts larger than 3 and 4, respectively 
(e.g., \citealt{perezgonzalez05}; \citealt{magnelli13}; \citealt{gruppioni13}) in the more sensitive bands (e.g., {\em Spitzer} $24\,\mu$m and {\em Herschel} $100\,\mu$m), while the {\em Herschel} longer wavelength instrument (SPIRE, observing at 250, 350 and $500\,\mu$m), did detect potentially higher redshift galaxies although they are often difficult to identify, due either to the large obscuration at optical 
wavelengths or to the very large beam-size of SPIRE ($\sim$18~arcsec at $250\,\mu$m; $\sim$35~arcsec at $500\,\mu$m)). In any case, the deepest cosmological surveys performed by 
{\it Herschel} at high-$z$ have detected only the most luminous galaxies ($L_{\rm IR}$$>$10$^{12}$~L$_{\odot}$ at $z$$\sim$3; \citealt{gruppioni13}, \citealt{magnelli13}). 
Previous IR space telescopes did not have enough sensitivity, while the upcoming James Webb Space Telescope (JWST), due to launch in 2018, will have a very narrow field of view (FoV), 
which will reflect in a very low mapping speed for surveys.

SPICA (e.g., \citealt{nakagawa12}; \citealt{sibthorpe15}; Roelfsema et al. 2017), an IR space observatory with a 2.5-m primary mirror cooled to about 6 K, and with a new generation of ultra-sensitive detector arrays, will offer the community a unique astronomical facility, covering the rich 12--$350\,\mu$m spectral range, capable of making deep and wide surveys to unprecedented depths in spectroscopy, photometry, and polarimetry.
The two instruments on board of SPICA, SAFARI, a joint European-Canadian-US contribution, and SMI from Japan, together provide several modes of operation with high resolution 
(HR, $R\simeq$28000) spectroscopy in the mid-IR (12--$18\,\mu$m) and low (LR, $R\simeq$150) to medium resolution (MR, $R$ up to $\simeq$11000) spectroscopy instantaneously over the full 
17 to $36\,\mu$m and 35 to $230\,\mu$m ranges, at a sensitivity of a few times 10$^{-20}$ Wm$^{-2}$ (5$\sigma$, 10 hr). 
In addition to the mid- (SMI) and far-IR (SAFARI) spectrographs, driving the main science objectives with SPICA (see Spinoglio et al. 2017 this issue, and Kaneda et al. 2017 submitted to PASA as part of this issue), 
the SMI/LR capability provides a 10$^{\prime}$$\times$12$^{\prime}$ slit viewer camera (SMI/CAM, a broadband imager centred at $34\,\mu$m, with a 30--$37\,\mu$m band), 
which is perfectly suited to perform wide-area surveys, while SAFARI delivers imaging polarimetry (SAFARI/POL at 100, 200 and $350\,\mu$m) and photometry-mapping in the far-IR 
(centred at 45, 72, 115 and $185\,\mu$m). 
In the SMI/LR spectral mapping mode, the multi-slit spectrometer and the camera are operated simultaneously, yielding multi-object spectra from 17 to $36\,\mu$m and $R=$5 deep imaging 
at $34\,\mu$m (see \citealt{kaneda17}).
The 30--$37\,\mu$m band will be very important for the detection of warm dust at high redshift, providing the ``sweet spot'' for unveiling the high-$z$ elusive/obscured AGN escaping 
detection at any other wavelengths.

For a detailed description of SPICA instrumentation and capabilities, and of its main scientific goals, 
we refer to Roelfsema et al. (2017, A\&A in preparation) and \citet{kaneda17}. 
\\ \\
This paper focuses mainly on the study of high-$z$ ($>$3) SF galaxies and AGN with photometric surveys in the mid-IR, and is organised as follows.  
In Section 2 we focus on the reasons why we need deep mid-IR photometric surveys; in Section 3 we discuss the main 
contribution of deep mid-/far-IR photometric surveys to our understanding of the SFRD and BHARD cosmic evolution up to the reionisation epoch. 
In Section 4 we present our concept for the best photometric 
survey strategy with an observatory like SPICA, with predictions of expected numbers of sources and 
redshift distributions, while in Sections 5 and 6 we discuss the importance of such surveys for the detection of highly obscured AGN and starburst galaxies, respectively. 
In Section 7 we present our conclusions.

Throughout this paper, we use a Chabrier (2003) initial mass function (IMF) and we adopt a $\Lambda$CDM cosmology with $H_{\rm 0}$\,=\,70~km~s$^{-1}$\,Mpc$^{-1}$, $\Omega_{\rm m}$\,=\,0.3, and $\Omega_{\rm \Lambda}\,=\,0.7$. 

\section{The need for a deep mid-IR photometric survey with SPICA} 
\label{midIR}
A deep and large-area photometric survey in the band covered by SMI/CAM on SPICA (30--$37\,\mu$m, centered at $34\,\mu$m) will be essential 
within the galaxy formation and evolution context, in particular for:
\begin{itemize}
\item[a)] unveiling the dust obscured population (either AGN or star-forming) up to high redshift (i.e., the reionisation epoch), providing a conclusive measure of the obscured 
SFRD and BHRD from $z$$=$0 to $z$$\simeq$6--7;
\item[b)] effectively detecting heavily obscured Compton-thick and elusive AGN missed by deep observations in the optical, UV and X-ray bands;
\item[c)] efficiently mapping large areas of the sky with the wide FOV, creating a blind catalogue for unbiased target selection for mid- and far-IR spectroscopic follow-up;
\item[d)] filling the wavelength gap between the currently existing and forthcoming facilities operating in the IR domain (e.g., between the JWST, 
0.6--$23\,\mu$m range, and ALMA, 350--$3500\,\mu$m).
\end{itemize}

\section{OBSCURED SFRD AND BHARD AT HIGH-$z$ with SPICA}
\label{sfrd_bhard}
The emission from galaxies and AGN at different wavelengths provides information about different physical processes within them; thus, to link the different physical processes 
together it is necessary to observe the same galaxies at many wavelengths.
Although many current, forthcoming and future facilities, including ALMA, the JWST, the Extremely Large Telescopes (ELTs) and the 
Advanced Telescope for High-ENergy Astrophysics (ATHENA), will dedicate a significant effort towards the study of the first luminous sources formed in the Universe, a 
wavelength gap in a band (mid-/far-IR) crucial for understanding the link between star-formation and BH accretion will remain. 

SPICA SMI/CAM, observing in photometric mode in the previously unexplored 30--$37\,\mu$m band, will be so sensitive 
that it will reach the confusion limit (estimated at $\sim$$9\,\mu$Jy at 5$\sigma$) in slightly more than 1 hour, thus enabling large areas of sky to be covered in relatively small 
amounts of time and detecting star-forming galaxies and obscured AGN out to $z$$\sim$6--7.
Producing statistically significant, confusion-limited surveys in the mid-IR will be a unique capability of SPICA. 
A multi-tiered photometric survey with SPICA-SMI/CAM will enable us to identify a mid-IR unbiased sample of galaxies and AGN over a 
wide range in redshifts, pushing the study of galaxy and AGN evolution all the way back to the reionisation epoch.

A SPICA photometric survey will contribute to our knowledge of the cosmic SFRD and BHARD at three different levels, depending on the availability of either spectroscopic or 
multi-wavelength ancillary data.
\begin{itemize}
\item[1)] At 0$<$$z$$<$3 high resolution spectroscopic observations of SMI/CAM-targeted sources with SAFARI will provide important indicators of the physics of the ISM -- particularly of the 
ionised gas -- within galaxies and AGN at the peak of the obscuration and SF/accretion epoch. 
Detailed spectroscopy of individual galaxies selected from wide-area surveys provides a unique suite of diagnostic tools accessible only in the IR.
These diagnostics, based on mid- and far-IR lines, are fundamental for characterising the physics of the ISM in dusty galaxies 
(see Spinoglio et al. 2017, this issue). 
IR spectroscopy will allow us both to measure the redshift of dusty sources and to link the different line properties to the IR luminosity, was well as the SFR and the AGN luminosity (if present), 
as described in section~\ref{irspec}. Moreover, IR spectroscopy will allow the detection of molecular gas outflows driven either by AGN or SF 
(see Gonz\'alez-Alfonso et al. 2017, this issue), and will also measure the amount of metals and dust within galaxies (see Fern\'andez-Ontiveros et al. 2017, this issue). 
\item[2)] At $z$$>$3 (and lower), simultaneous observations 
with the SMI/LR (R$\simeq$50--120) will detect features in the mid-IR range (i.e., polycyclic aromatic hydrocarbons, PAHs: the 3-$\mu$m rest-frame feature will be in the 
SMI range up to $z$$\simeq$11, the 9.7-$\mu$m silicate absorption to $z$$\simeq$3) of normal SF galaxies, which represent key tracers of SFR and AGN accretion and can be used to determine redshift. 
\citet{Kaneda17} estimate that at redshifts greater than $\sim$4, the PAH features should be detectable only in extremely luminous sources (ultra/hyper Luminous IR Galaxies: LIRGs/HyLIRGs), 
with $L_{\rm IR} \geq 10^{12.5-13}$ L$_{\odot}$ (although the PAH properties and strengths at high redshifts might be different from what we expect based on lower-$z$ observation).
\item[3)] For galaxies either at higher redshifts (greater than $\sim$4) or having lower luminosity (not as extreme as ULIRG/HyLIRGs, i.e., with $L_{\rm IR}$$<$10$^{12}$ L$_\odot$), not detected in high/low-resolution spectroscopy by either SAFARI or SMI/LR, ancillary data would be necessary, in order to provide identifications, measure the redshifts and quantify the SFRs and/or AGN accretion rates. For this reason, the deep SPICA photometric surveys will be performed on extensively observed fields, where large multi-wavelength coverage will be available (see, e.g., the {\em Herschel} Extragalactic Legacy Project HELP, \citealt{vaccari16}, providing homogenised multi-wavelength datasets on all the {\em Herschel} extragalactic wide-area fields). 
Follow-up observations with SAFARI in photometric mode at $45\,\mu$m will help shaping the SEDs, providing a measure at longer wavelengths than the detection one.
For the SPICA sources unidentified by any other means (which will be particularly interesting, since they will 
open up a window to a potentially unknown population), specific follow-up campaigns will be needed, with, e.g., ALMA, ELTs and other facilities that will be operational at the time 
SPICA is expected to fly. As shown in Sections~\ref{survey}, \ref{AGN} and \ref{SB}, ALMA and ELTs will be able to identify all the sources detected by SPICA-SMI/CAM at 34\,$\mu$m, 
up to very high redshifts ($z$$\gtrsim$8). 
\end{itemize}


\subsection{Calibrate the $z$$<$3--4 SED-fitting physical quantities through IR lines}
\label{irspec}
As extensively discussed by Spinoglio et al. (2017, this issue), since the emission-line intensities and ratios in the mid- and far-IR domain do not suffer from dust extinction
like the optical and UV emission lines, they provide unique information on the physical conditions (i.e., electron
density and temperature, degree of ionisation and excitation and chemical composition) of the gas within the dust-obscured regions 
of galaxies with intense SF activity or surrounding an AGN
(\citealt{spinoglio92}; \citealt{rubin94}; \citealt{panuzzo03}). 
To properly quantify the contribution of AGN and SF from young massive stars
to the overall energy budget in large samples of dusty obscured galaxies, mid- and far-IR lines can be used as proxies of 
the accretion rate (e.g., [O~IV]25.9$\mu$m and [Ne~V]14.3, 24.3$\mu$m) and the SFR
(e.g., [Ne~II]12.8$\mu$m, [S~III]18.7, 33.5$\mu$m  and [O~III]52, 88$\mu$m), since the luminosity of the mid-/far-IR lines has been found to trace and strongly correlate 
with either the SF or the AGN luminosity in galaxies (e.g., \citealt{spinoglio12}; \citealt{bonato14a, bonato14b}; \citealt{delooze14}; \citealt{gruppioni16}). 
These relations, derived only in the local Universe for a limited number of objects, depend on the relative AGN contribution to the IR luminosity. 
SPICA, measuring spectra for thousands of galaxies up to $z$$\simeq$ 3--4, will allow us to derive similar relations for statistically significant samples of sources 
spanning a wide range of luminosities, allowing us to test their evolution and understand the physical processes at play. 

\begin{figure}
\begin{center}
\includegraphics[width=\columnwidth]{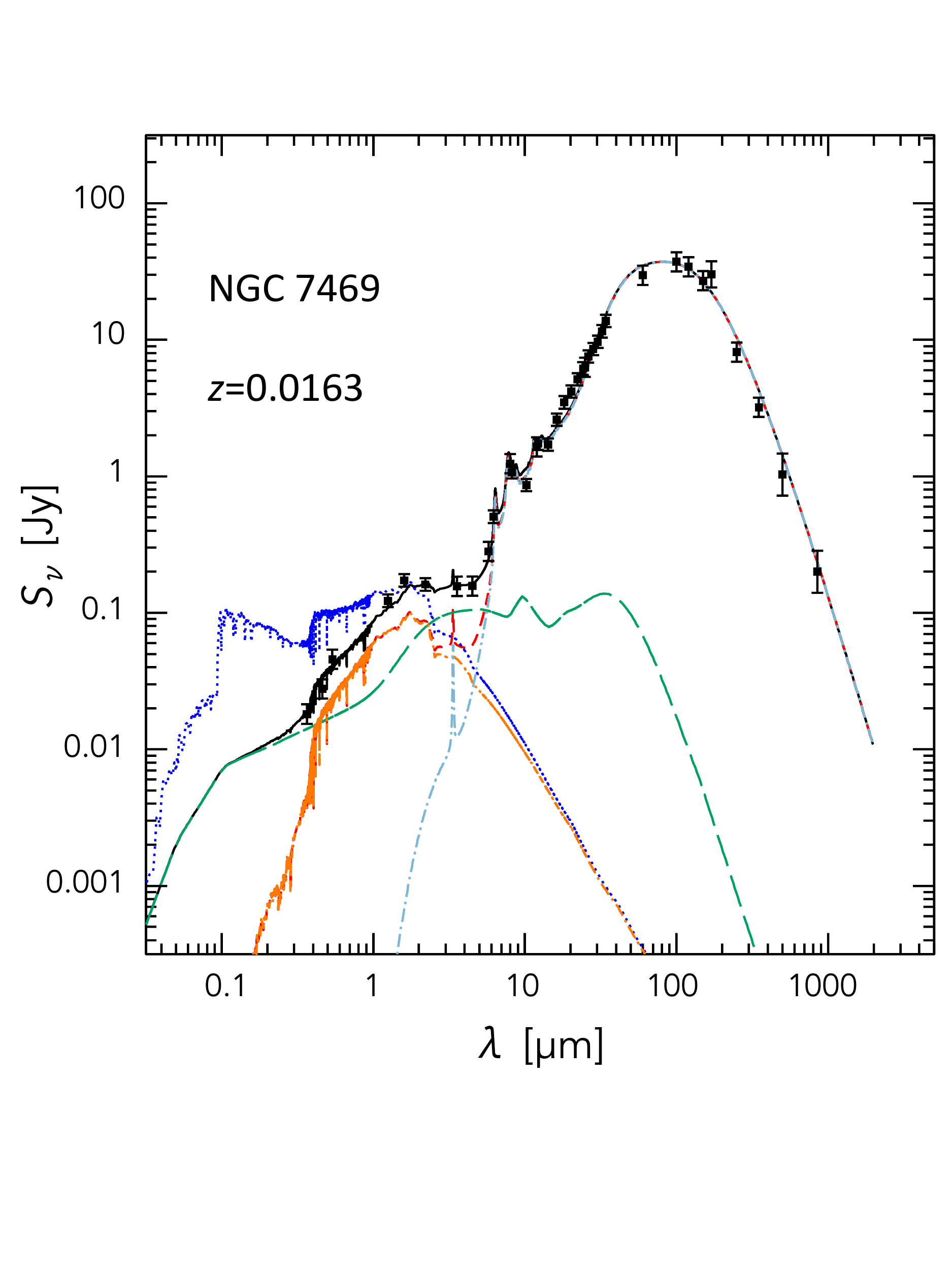}
\caption{Example of an observed SED decomposed into stellar, AGN and star-formation components, using the technique developed by \citet{berta13}. The black filled circles with 
error bars are the photometric data relative to the optically classified Seyfert 1 NGC7469. The blue dotted line shows the unabsorbed stellar component, the red dashed line shows the combination of extinguished stars and dust IR emission, while the long-dashed green line shows the dusty torus emission. The pale-blue dot-dashed line shows the dust re-emission, while the black solid line
is the sum of all components (total emission).}
\label{sed}
\end{center}
\end{figure}

In this context, \citet{gruppioni16} have obtained new estimates of AGN accretion and SF luminosity for local Seyfert galaxies from the 12-$\mu$m sample of \citet{rush93}, 
by performing a detailed broadband SED decomposition including the emission of stars, dust heated by SF, and a possible AGN dusty torus 
(see Figure \ref{sed} for an example of an SED consistently decomposed into its main building blocks). 
\begin{figure}
\begin{center}
\includegraphics[width=\columnwidth]{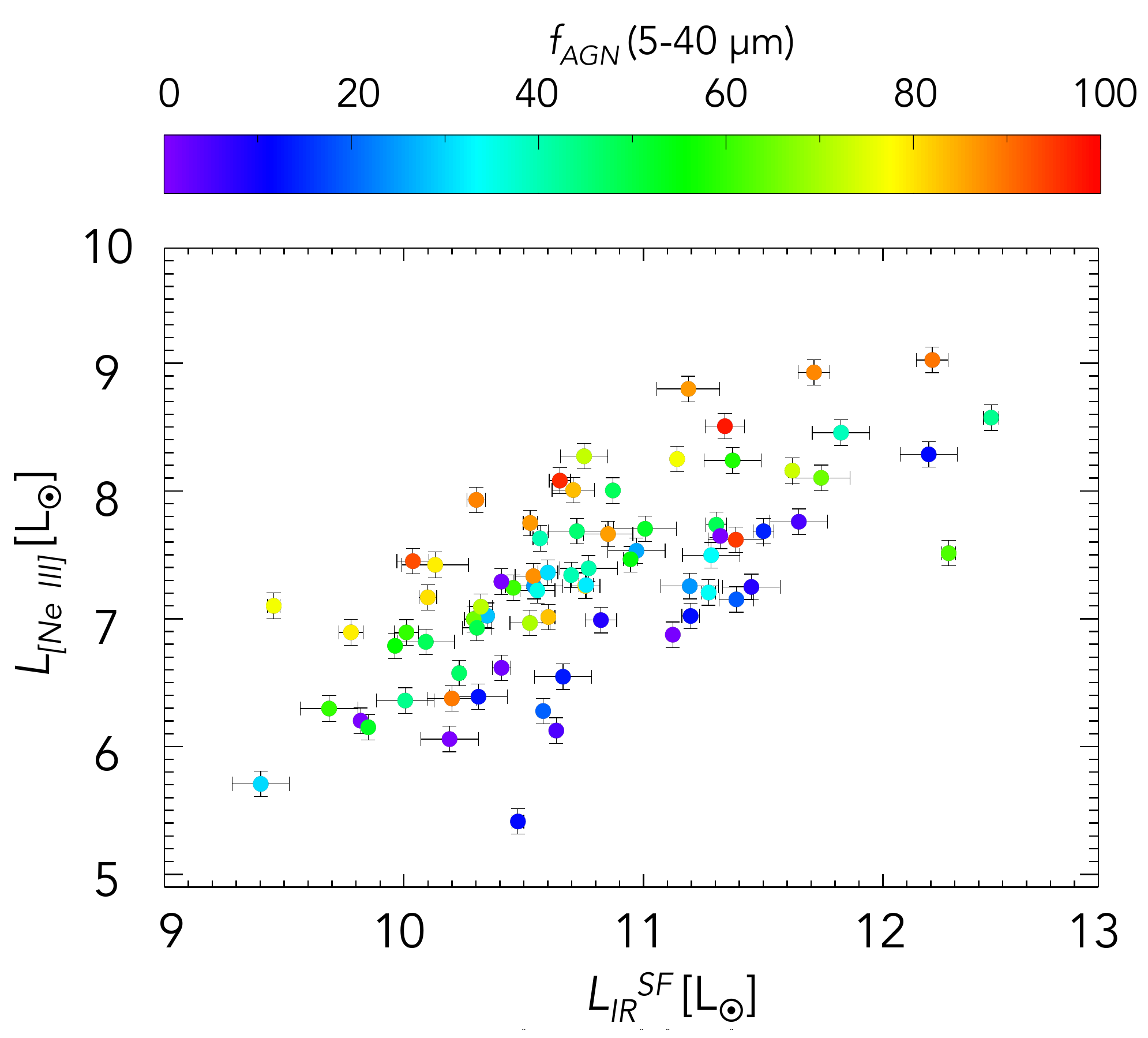}
\caption{Luminosity of the [Ne~III]~15.6$\mu$m line as a function of the 8--$1000\,\mu$m luminosity due to SF ($L_{\rm IR}^{\rm SF}$), as derived from the 
SED decomposition analysis (e.g., see Figure~\ref{sed}, pale blue dot-dashed line) of the 12-$\mu$m sample of local galaxies performed by \citet{gruppioni16}. 
The different colours of the symbols represent the different AGN fractions to the 5--$40\,\mu$m luminosity (i.e., $f_{AGN}(5-40~\mu m)$ in the top colour-bar).}
\label{lneiii_lsf}
\end{center}
\end{figure}
The IR luminosity due to SF, and the intrinsic AGN bolometric luminosity derived through SED-fitting, have been shown to strongly correlate with the IR line luminosity, with the scatter 
mainly due to different relative AGN contributions for different galaxies in the sample (see Figure \ref{lneiii_lsf} for an example of the correlation between the [Ne~III]~15.6$\mu$m line luminosity and the IR luminosity due to SF, with different colours highlighting different AGN fractions). 
By extrapolating the local derivations of {\em Herschel}-detected line emission and SF or AGN luminosities to higher redshifts, it was possible to obtain mid- and far-IR line luminosity functions (\citealt{gruppioni16}). These estimates, used in combination with IR galaxy and AGN evolutionary models, are now useful for making predictions of what we expect to observe with SPICA
(in terms of emission lines and numbers of objects detectable in the different lines) up to $z$$\sim$3--4 and beyond (see Spinoglio et al. 2017, this issue). 

SPICA will thus extend these studies to a redshift of 3 to 4, directly linking the
physical quantities obtainable from photometric observations (e.g., SED fitting) with emission line (ISM) properties, in fainter and higher redshift galaxies than the local 
Seyferts analysed by \citet{gruppioni16}. These relations, whose evolution with redshift will be studied in detail, will be extremely useful in calibrating the physical quantities derived through 
SED-fitting using IR lines up to $z$$\sim$3. Then we will use these calibrations at higher $z$, to estimate the physical properties of the ISM and the nature of the heating source, for 
galaxies out of reach of SPICA spectroscopy (due to sensitivity limits of the spectrographs). 

\subsection{Obscured SFRD and BHARD evolution from their peak to reionisation ($z$$>$4)}
\label{photo}
The exquisite photometric capabilities of SPICA will make it possible to resolve the existing discrepancy in the contribution of dust-obscured sources to the $z$$>$4 SFRD (see Figure \ref{sfrdz}).
This is also likely to impact on the estimate of the relative contribution of galaxies and AGN to the reionisation (see, e.g., \citealt{fontanot12, fontanot14}).
As mentioned in the previous section, galaxies at $z$$>$3--4, unless strongly lensed, will be hardly detectable in spectroscopic mode by SPICA. 
High-$z$ sources could therefore be detected only in photometric mode,   
using the SMI/CAM, which will be sufficiently sensitive to easily reach the confusion limit 
for a 2.5-m telescope (estimated 9\,$\mu$Jy, 5$\sigma$; see section~\ref{sec_confusion}).

\begin{figure}
\begin{center}
\includegraphics[width=\columnwidth]{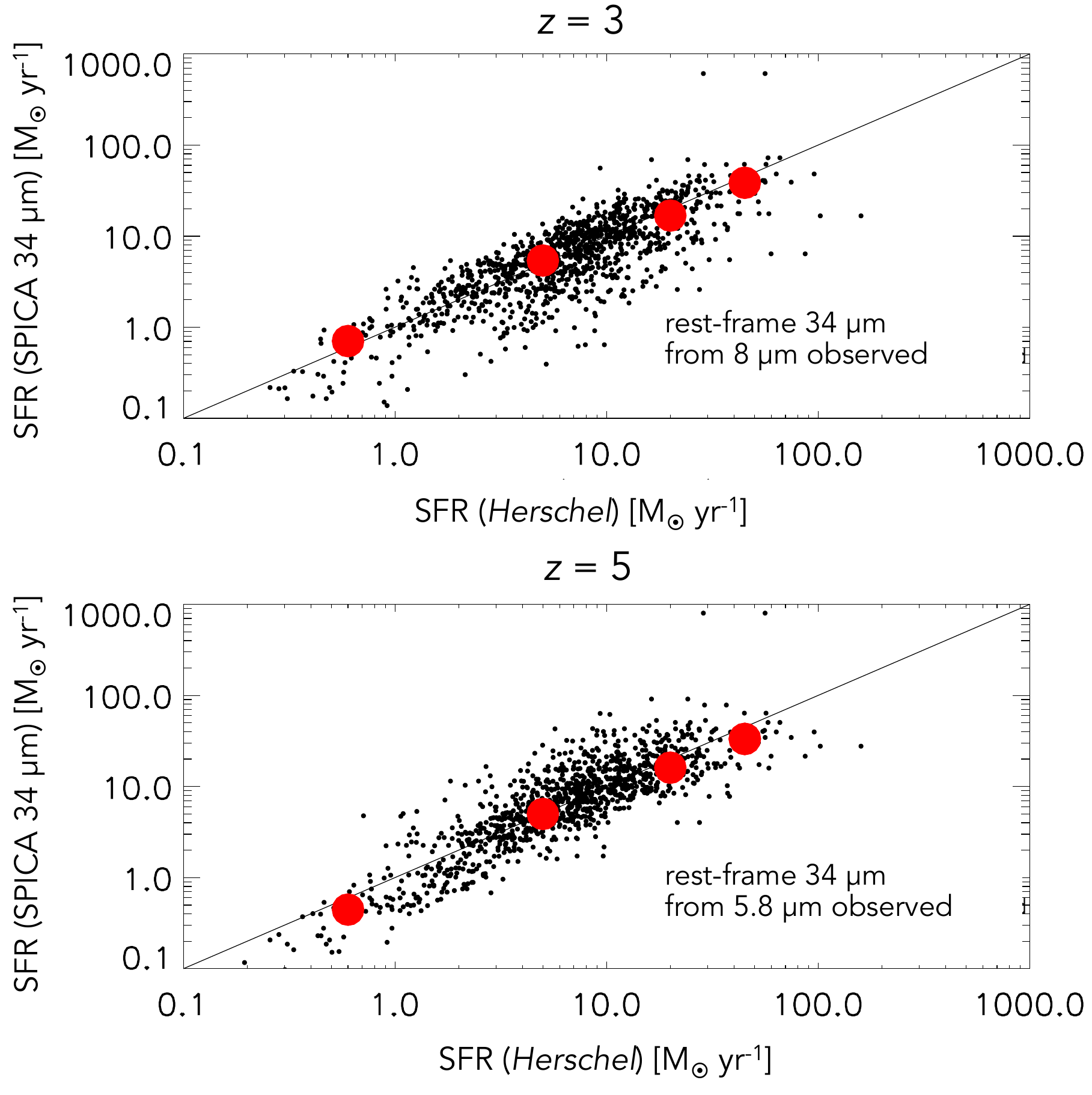}
\caption{Examples of how the photometric SPICA survey at $34\,\mu$m could be able to recover the SFR of star-forming galaxies
at $z$$=$3 and $z$$=$5. We compare the SFR (effectively the total IR luminosity) extrapolated from the observed 34-$\mu$m
flux density at $z$$=$3 ({\it top panel}) to that computed when including photometric data from {\it Herschel}.
To perform this test we used a sample of local galaxies ($z$$<$0.3) from the COSMOS survey (\citealt{laigle16}),
where the observed 8-$\mu$m IRAC fluxes mimic the rest-frame emission of the 34-$\mu$m filter at $z$$=$3.
Similarly, we adopted the 5.8-$\mu$m IRAC fluxes ({\it bottom panel}) to probe the 34-$\mu$m filter at $z$$=$5. 
The red filled circles show the median values in bins of SFR.}
\label{SFR_spica}
\end{center}
\end{figure}

The validity of the 34-$\mu$m (continuum) SFR indicator alone (although a redshift estimate/measure is needed) has been tested at $z$$=$3--5 by extrapolating the total IR
luminosities (converted to SFRs) estimated from the observed 34-$\mu$m luminosity. In fact, the $34\,\mu$m observed band at $z$$\simeq$3 and 5 samples the 
8 and 5.8-$\mu$m continuum rest-frame respectively.
A measure can be obtained by tuning the SED templates 
as calibrated by {\em Spitzer} and {\em Herschel} (\citealt{magdis12}), with a spread of about a factor of 1.8 (comparable to the intrinsic scatter of the SFR-stellar 
mass relation). The relation (shown in Figure \ref{SFR_spica}) is very promising, because it could be extremely useful to derive the SFR
of the 34-$\mu$m selected sources without far-IR counterparts (e.g., too faint to be detected by SAFARI above confusion), and thus obtain their IR 
luminosities without the need to integrate over their SEDs. To perform this test we used the SPLASH/IRAC data for local galaxies ($z$$<$0.3) from the COSMOS survey (\citealt{laigle16}), 
where the observed 8-$\mu$m and 5.8-$\mu$m IRAC fluxes mimic the rest-frame emission of the 34-$\mu$m filter at $z$$=$3 and 5, respectively. 
Although the relations are favourable, they are based on the assumption that the properties of galaxies at $z$$\geq$4 are similar to those at $z$$\sim$2. 
We do, however, know that  the typical SED of ULIRGs at $z$$\sim$2 is different from that at $z$$\sim$0 and therefore this result has to be treated with caution.
For the high-$z$ galaxies, follow-up with ALMA, IRAM or single-dish sub-mm observatories could help to calibrate these relations, providing better estimates of the total IR luminosity. 

\section{A PHOTOMETRIC SURVEY WITH SPICA-SMI/CAM}
\label{photo_survey}
While we expect hundreds of high-$z$, IR-selected sources to be available for follow up by the time
SPICA is expected to fly (e.g., built up from surveys with {\em Herschel}, ALMA, SPT, JWST, Euclid
and WFIRST), SPICA with SMI will be uniquely capable of discovering new galaxies. Unlike SPICA, ALMA and the JWST are not designed to map large regions of the 
sky but will be able to make detailed observations of the objects uncovered by SPICA.
Multi-tiered photometric surveys with SPICA-SMI in the newly available 30--$37\,\mu$m band, will be used to select an unbiased sample of galaxies and AGN over a wide range of redshifts.  
These observations can be followed up by SAFARI in both spectroscopic and photometric mode, pushing 
the study of galaxy and AGN evolution up to $z$$\sim$6, and testing the claim that at $z$$>$4 dust attenuation is negligible.
Indeed, very recently \citet{laporte17} found a (gravitationally lensed) star-forming galaxy at $z$$\simeq$8, with an estimated dust mass (through ALMA data) 
of 6$\times$10$^6$~M$_\odot$. Similar (or larger) dust masses, if found in the majority of high-$z$ galaxies, will have important implications
for our understanding of galaxy (and dust) formation and evolution.

\subsection{Estimated source confusion for a 2.5-m mirror telescope}
\label{sec_confusion}
\begin{figure}
\begin{center}
\includegraphics[width=\columnwidth]{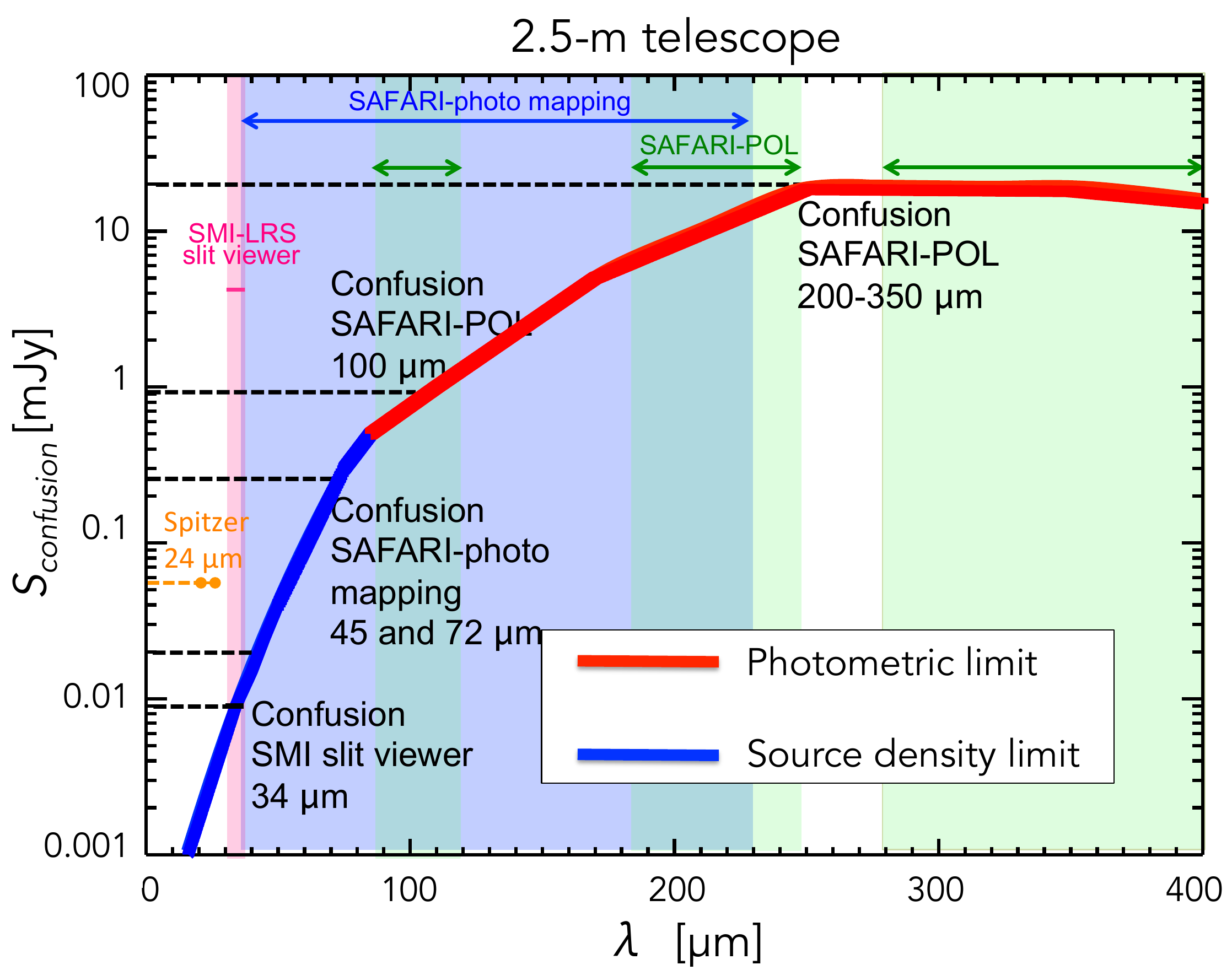}
\caption{Confusion limit as a function of wavelength for a diffraction-limited 2.5-m telescope. The limits of the SMI/CAM (30--$37\,\mu$m) and of SAFARI 
(/POL at 100, 200 and $350\,\mu$m, and in photometry mapping
at 45 and $72\,\mu$m) are shown as horizontal dashed lines, while the pink, blue and green shaded areas show the wavelength ranges covered by 
SMI/CAM, SAFARI photometric mapping and SAFARI/POL, respectively. The blue part of the curve is determined by the source density criterion, while the 
red part is defined by the photometric criterion. For comparison, the confusion limit reached by {\em Spitzer} (with a 0.85-m mirror) at $24\,\mu$ (from \citealt{dole04})
is shown as orange horizontal line.}
\label{confusion}
\end{center}
\end{figure}
To estimate the confusion limit expected for a given sized telescope (2.5~m in our case), it is necessary to extrapolate the extragalactic source counts down to very faint flux densities and to consider 
the diffraction-limited beam size of the telescope.
For source counts, we have considered the models of \citet{pozzi15}, which are able to reproduce the latest {\em Herschel}~survey data, extrapolated to fainter fluxes. 
We have followed \citet{dole03} to determine the confusion 
limits that will affect photometric surveys taken with a given angular resolution. 
According to \citet{dole03}, there are two sources of confusion:
1) the photometric limit (i.e., the noise produced by sources fainter than the detection threshold), where the photometric criterion is the requirement that sources are detected with a S/N$>$5; and 2) the source density limit (i.e., depending on the fraction of ``blended'' sources). For the source density criterion, we require that $<$30\% of the sources are close enough not to be separable, i.e., are within 0.8$\times$FWHM of another source). 
We use the higher of these values at a given wavelength to set the confusion limit to which we will integrate to in the SPICA maps. 
As shown in Figure \ref{confusion}, at $\lambda$$\leq$70--$80\,\mu$m the main limitation for deep photometric surveys is the fraction of blended sources (blue curve), while at $\lambda$$>$70--$80\,\mu$m the photometric limit dominates (red curve).

While the confusion limit is roughly constant above $200\,\mu$m, the combination of the source number density variation with wavelength and the reduction in beam size at as wavelength 
decreases means that the confusion flux density decreases dramatically at shorter wavelengths. 
This will allow SPICA, given its high sensitivity, to go much deeper at 35 and $45\,\mu$m without hitting confusion, than did PACS at $100\,\mu$m. 
The estimated 5$\sigma$ confusion of SMI/CAM (30--$37\,\mu$m) is $\sim$$9\,\mu$Jy, while the SAFARI confusion 
is $\sim$0.02, 0.25 and 1\,mJy at 45, 72 and $100\,\mu$m, respectively, and 18 mJy at $>$$200\,\mu$m 
(the latter values are confirmed also by the {\em Herschel} results of \citealt{magnelli13} in the H-GOODS Survey and of \citealt{nguyen10} in the HerMES survey, reaching confusion at 
0.8 and 19\,mJy at 100 and $250\,\mu$m, respectively).
Given the estimated confusion limits, it will be possible to use the $45\,\mu$m channel of SAFARI (in photometric mode, i.e., used by binning the spectra) to follow-up targeted 
SMI sources (the FOV of SAFARI in photometric mapping is only 1$^{\prime}$$\times$1$^{\prime}$) out to relatively high redshifts, going 2--3$\times$ below confusion.
We note that the confusion estimates depend on source counts extrapolations, and that different criteria to derive confusion can provide different values. As an example,
if we follow the method described in  \citet{franceschini89}, we'd find limits 2--3 times lower than our estimated values at wavelengths 
$<$$60\,\mu$m (i.e., where we are dominated by the source density criterion).  
In fact the only way to determine the actual level of confusion (and thereby constrain galaxy evolution and source count models) is to make the measurements -- for which we require 
an instrument like SPICA.

\subsection{Survey strategy}
\label{survey}
The optimal survey structure envisaged to perform evolutionary studies consists of multiple layers of different sizes and depths (i.e., the classical ``wedding-cake''). 
Here we present a plan for three surveys with SMI/CAM in the 30--$37\,\mu$m band.
\begin{itemize}
\item[1.] {\bf Ultra-deep survey (UDS)}: a sub-confusion survey (down to 5$\sigma$$\simeq$$3\,\mu$Jy -- i.e., 3 times lower than the estimated confusion) of 
0.2 deg$^{2}$, to examine the confusion noise itself. 
This could be done in just a few fields with extensive multi-wavelength coverage (e.g., CANDELS fields). According to the expected nominal sensitivities, 
the estimated time (only observational, excluding overheads) needed for such a survey is 
around 100~hours (six pointings).
\item[2.] {\bf Deep survey (DS)}: A survey of  $\sim$1 deg$^{2}$ in area down to the 5$\sigma$ confusion limit of $9\,\mu$Jy, which will be reached in 1.75\,min exposure per pointing. 
Since 36 pointings are needed to cover the desired area, a total of 64 hours will be necessary to perform the survey (for comparison, the Design Reference Mission, DRM, for JWST including a deep -- 
$16\,\mu$Jy, 5$\sigma$ -- survey at $21\,\mu$m with MIRI, over a field of 10$^{\prime}$$\times$9$^{\prime}$,  will take 160~hours to complete).
The SPICA DS should be performed on an extensively observed area of the sky of similar size (e.g., COSMOS), to take advantage of the available multi-wavelength coverage.
\item[3.] {\bf Shallow survey (SS)}: 600~deg$^2$ (to study the bright-end of the LF) to 5$\sigma$ about 0.2\,mJy. The estimated time need without instrument overhead is about 78~hours (21600 pointings to cover the whole area). 
A wide-area multi-wavelength surveyed region of the sky to be covered by the SS, could be part of the GAMA {\em Herschel}-ATLAS survey (\citealt{eales10}), or the Euclid Deep Survey fields.
\end{itemize}
If we could ignore the limited lifetime of SPICA and the need to share time with other projects, we could imagine a very large area survey, covering about 3000\,deg$^2$ with SMI/CAM (performed on fields like the deep LSST or GAMA fields). Such a wide-area survey, considered as an example of the excellent SPICA-SMI/CAM performance as a photometer, would take about 390~hours
to reach a depth of 0.2\,mJy, and would be instead time-prohibitive for JWST-MIRI.
\begin{figure}
\includegraphics[width=8.5cm]{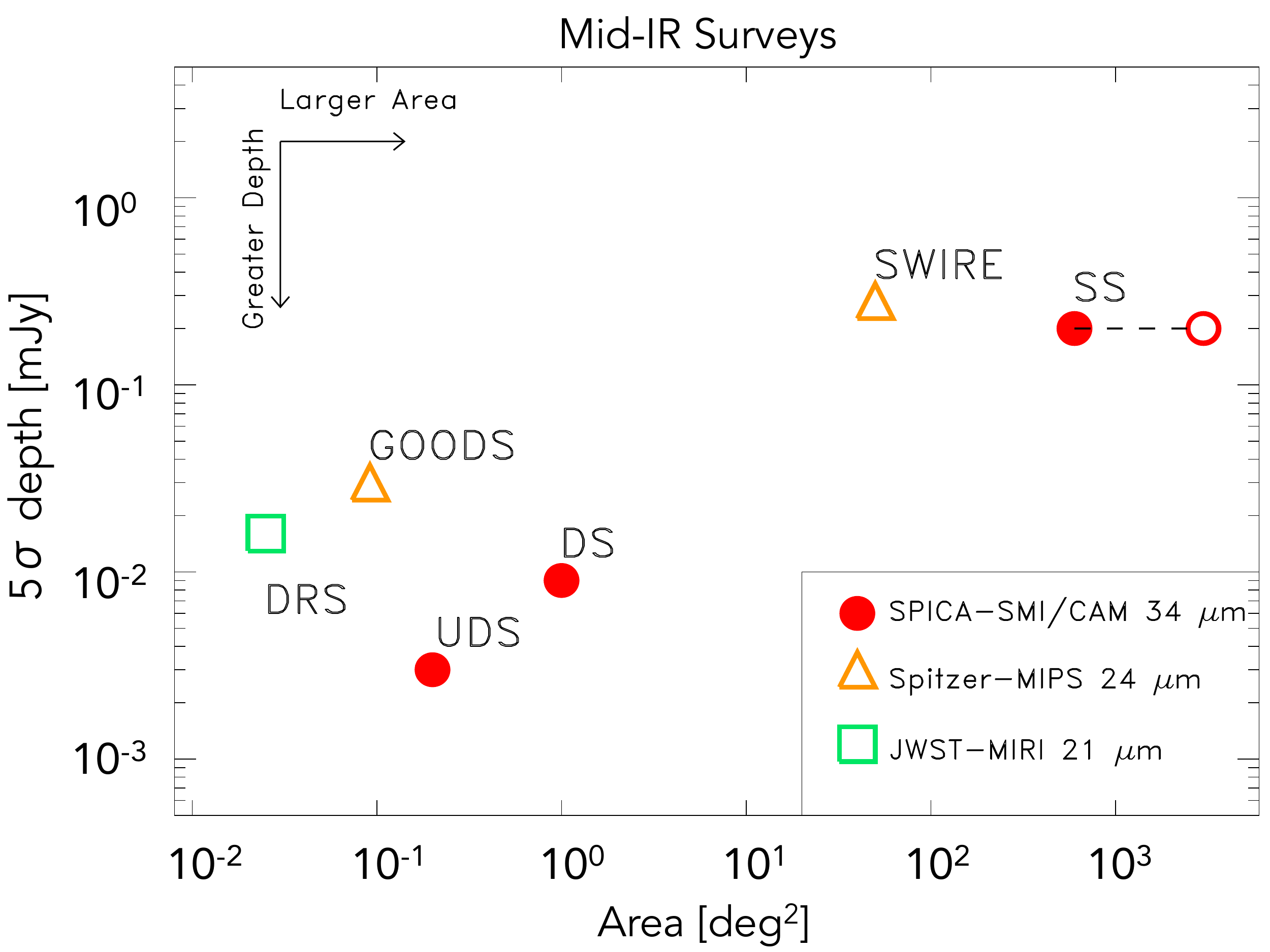}
\caption{Depth versus survey area for the planned extragalactic photometric surveys with SMI/CAM at $34\,\mu$m (red filled circles), compared to the values reached by the surveys performed 
with {\em Spitzer}-MIPS at $24\,\mu$m (the deepest, GOODS, and the largest, SWIRE; orange open triangles), and what will be reached by the Design Reference Mission survey planned 
with JWST-MIRI at $21\,\mu$m (green open square).
The red open circle represents the example of an extremely wide area survey (3000 deg$^2$) that could potentially be performed with SMI/CAM (in 390 hours).}
\label{fig_area_sens}
\end{figure}

In Figure~\ref{fig_area_sens} we show the depth reached versus area covered for the planned extragalactic photometric surveys with SMI/CAM at $34\,\mu$m (red filled circles) described above
(the UDS, DS and SS, including the example of an extremely wide area survey). compared to those performed with {\em Spitzer}-MIPS 
at $24\,\mu$m (the deepest, GOODS, and the largest, SWIRE; orange open triangles), and to the Design Reference Mission survey planned with JWST-MIRI at $21\,\mu$m (green open square).
From the figure the great improvement of SPICA is immediately clear with respect to {\em Spitzer} in both sensitivity (i.e., the deepest MIPS 24-$\mu$m survey, GOODS, was more than a factor of 
10 and more than a factor of 3 shallower than the planned UDS and DS, respectively) and area (i.e., the widest MIPS 24-$\mu$m survey, SWIRE, covered over 10 times less area than the planned SS).
Even not considering the much better speed efficiency of SPICA with respect to JWST in mapping the sky, the planned DRM survey with JWST-MIRI at $21\,\mu$m will be 5--6 times shallower 
than the SMI/CAM UDS, and will cover only about the 12\% of the area.

\begin{figure*}
\begin{center}
\includegraphics[width=17cm]{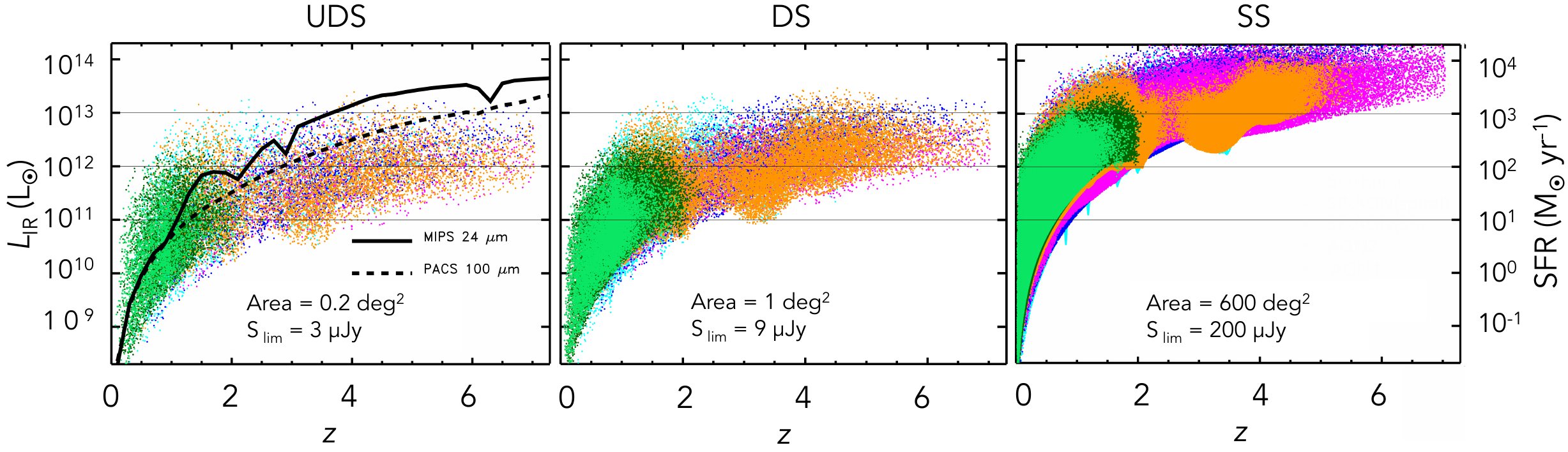}
\caption{Expected total IR luminosity (SFR) as a function of redshift for the three photometric surveys planned with SPICA-SMI/CAM ({\em left}: UDS to $3\,\mu$Jy; {\em middle}: DS to $9\,\mu$Jy; {\em right}: SS to 0.2\,mJy). The different colours of the points represent the behaviour of different SED-types: cyan, {\tt starburst}; green, {\tt spiral}; dark-green, {\tt SF-AGN(spiral)}; orange, {\tt SF-AGN(SB)}; magenta, {\tt AGN2}; and blue, {\tt AGN1} (see \citealt{gruppioni13} for details and the text for a brief description of these populations). 
For comparison, the limiting IR luminosity corresponding to the fluxes reached by the deepest surveys with {\em Spitzer} (MIPS $24\,\mu$m) and {\em Herschel} (PACS $100\,\mu$m) are shown 
in the {\it left panel} as black solid and dashed lines respectively.
The horizontal lines mark the LIRG, ULIRG and HyLIRG limits.}
\label{lir_z_smi}
\end{center}
\end{figure*}
\begin{figure*}
\begin{center}
\includegraphics[width=17cm]{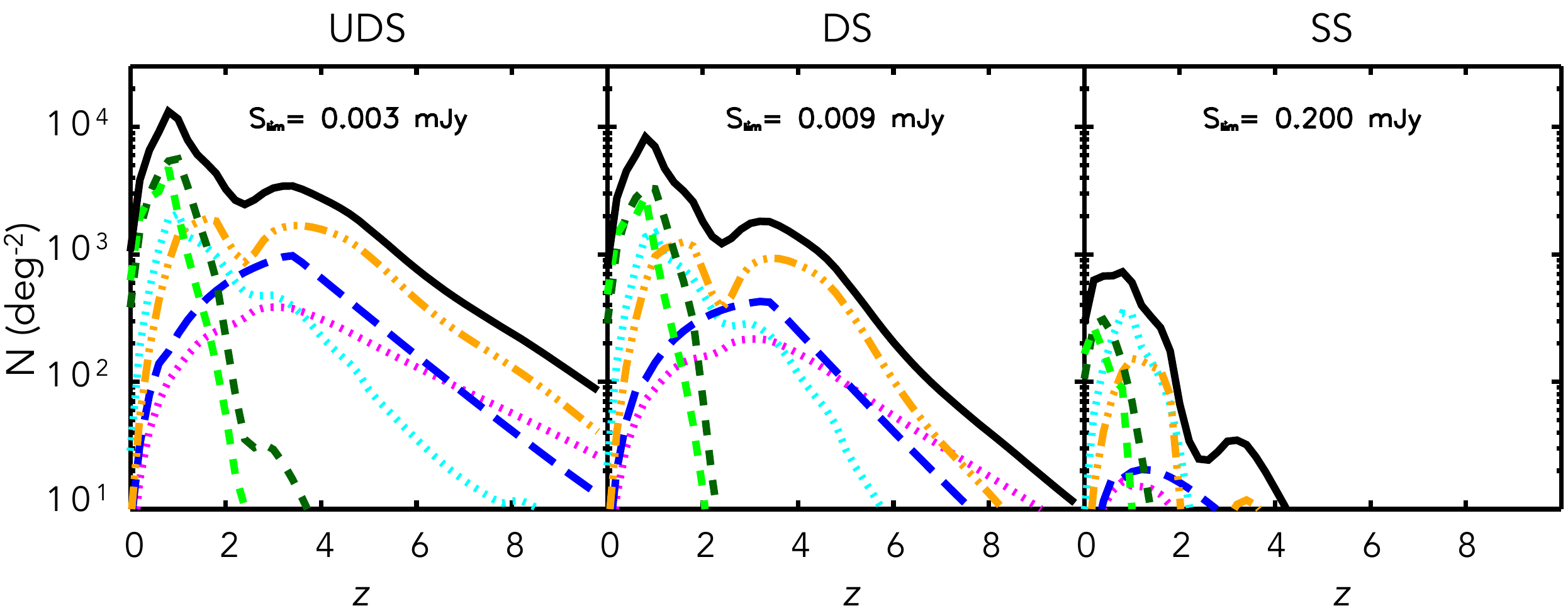}
\caption{Expected redshift distributions for the SMI photometric surveys (per square degree). The different lines and colours correspond to the different IR galaxy and AGN populations,
of Figure~\ref{lir_z_smi} (see caption and text description). The black solid line is the total redshift distribution, obtained by summing all the different populations.
The estimates are based on the {\em Herschel} LF evolution found by \citet{gruppioni13}, as revised by \citet{pozzi15}.}
\label{zdistr_smi}
\end{center}
\end{figure*}

The SPICA photometric surveys could be followed up by photometric mapping with SAFARI at 45, 72, 100, 250 and $350\,\mu$m. Deep surveys at these wavelengths are confusion-limited, with confusion 
corresponding to significantly higher fluxes ($40\,\mu$Jy at $45\,\mu$m, 1\,mJy at $100\,\mu$m, at 5$\sigma$) than the SMI ones, although, knowing the position of the 
34-$\mu$m sources, we can reliably extract flux down to lower limits (e.g., to 3$\sigma$). SAFARI, with its 1$^{\prime}$$\times$1$^{\prime}$ FOV, will be suitable for carrying out 
points source photometry on interesting sources selected from the SMI/CAM survey.
\begin{table}
\caption{Expected No. of sources in the SMI 34-$\mu$m survey.}
\begin{center}
\begin{tabular}{cccc}
\hline\hline
$z$ &  No. UDS & No. DS & No. SS \\
       &  Tot (AGN) & Tot (AGN) & Tot (AGN) \\
\hline%
 0--1  & 6.3 (3.0)$\times$10$^{3}$  &  2.1 (0.9)$\times$10$^{4}$ &1.7 (0.8)$\times$10$^{6}$\\
 1--2  &  7.3 (4.9)$\times$10$^{3}$ &  2.2 (1.5)$\times$10$^{4}$&1.4 (0.5)$\times$10$^{6}$\\ 
 2--3  & 2.7 (2.2)$\times$10$^{3}$ & 7.1 (5.4)$\times$10$^{3}$  & 1.4 (0.8)$\times$10$^{5}$\\ 
 3--4  & 3.2 (2.9)$\times$10$^{3}$ &  8.4 (7.3)$\times$10$^{3}$ & 9.0 (7.2)$\times$10$^{4}$\\ 
 4--5  & 2.3 (2.1)$\times$10$^{3}$ &  5.2 (4.9)$\times$10$^{3}$& 2.5 (2.4)$\times$10$^{4}$\\ 
 5--6  & 1.2 (1.1)$\times$10$^{3}$  &  1.9 (1.9)$\times$10$^{3}$& 4.1 (4.1)$\times$10$^{3}$\\ 
 6--7  &  5.6 (5.5)$\times$10$^{2}$ &  5.7 (5.7)$\times$10$^{2}$& 9.0 (9.0)$\times$10$^{2}$\\ 
 $>$7  & 1.2 (1.2)$\times$10$^{1}$ &  8.0 (8.0)$\times$10$^{0}$& 1.8 (1.8)$\times$10$^{1}$\\
\hline\hline
\end{tabular}
\end{center}
\label{tab_num_z}
\end{table}

In term of evolutionary studies, to investigate the evolution of IR galaxies and AGN through their LF, the UDS will be crucial for sampling the faint end of the LF by detecting low luminosity IR sources at high-redshifts 
(i.e., $L_{\rm IR}$$\simeq$10$^{11-11.5}$ L$_{\odot}$ to $z$$=$5--6, $L_{\rm IR}$$\simeq$10$^{12}$ L$_{\odot}$ to $z$$\gtrsim$6), while the DS will sample the knee of the IR LF, with a statistically significant number of galaxies. 
The SS, being unique in covering such a large area of the sky in IR, is expected to detect numerous bright IR sources to shape the bright-end of the LF with a very large statistics.
In Figure~\ref{lir_z_smi} we show a simulation of the 8--$1000\,\mu$m integrated luminosity, i.e., $L_{\rm IR}$, (and the corresponding SFR obtained through the \citealt{kennicutt98} relation scaled for a Chabrier IMF) and the redshift distribution expected for the three planned surveys (left: UDS, middle: DS, right: SS). 
The different colours represent the evolution of the different SED-types, as defined and studied by \citet{gruppioni13} and \citealt{pozzi15}:\\ 
${\bullet}$ normal {\tt spiral} galaxies (green);\\
${\bullet}$ pure {\tt starburst} sources (i.e., showing no signs of hosting an AGN, cyan);\\
${\bullet}$ star-forming galaxies hosting an AGN, which can be either low-luminosity inside a spiral-like galaxy ({\tt SF-AGN(spiral)}, dark-green), or obscured within a starburst galaxy ({\tt SF-AGN(SB)}, orange);\\
${\bullet}$ AGN-dominated sources, either optically unobscured ({\tt AGN1}, blue) or obscured ({\tt AGN2}, magenta). \\
Note that the distributions correspond to the
effective area of the planned surveys: in the wide area, shallow survey the number of expected sources is far larger than in the deeper surveys, with a significant tail also at
high redshift ($>$4--5), although it is mostly composed of extremely bright AGN-dominated sources ($\gtrsim$10$^{13}$~L$_\odot$). On the other hand, with the UDS we expect to detect galaxies
with $L_{\rm IR}$$\lesssim$10$^{10}$~L$_\odot$ up to $z$$\sim$2, and $L_{\rm IR}$$\lesssim$10$^{11}$~L$_\odot$ up to $z$$\sim$4.

The estimated redshift distributions (per unit area, i.e. in deg$^{-2}$) that could be obtained 
with the three SMI surveys are shown in Figure~\ref{zdistr_smi}.
These have been derived by considering the {\em Herschel} source evolution for the different IR populations described above, as found by \citealt{gruppioni13} and further modelled to 
higher redshifts by \citealt{pozzi15}, using a backward phenomenological approach, combined with spectrophotometric evolution of dust and SED evolution. 
Note that the estimates here have been extrapolated to $z$$\simeq$10 by considering the evolutions 
derived by \citet{pozzi15} up to $z$$\simeq$5. The distributions are shown per unit area, to give the idea of the different numbers and relative population contributions 
obtainable at different redshifts to different flux depths; these need to be scaled for the real area of each survey to derive the expected numbers in each of the 
planned configurations (in Table~\ref{tab_num_z} we present the effective numbers within the planned areas, while Figure~\ref{lir_z_smi}
shows a visual comparison between the surveys in terms of source density and depth). Note that the numbers given in Table~\ref{tab_num_z} are simply for redshift intervals
regardless of luminosity; even if the three surveys appear similar in terms of source numbers, one must consider also the reachable
luminosities when planning a survey. The different luminosities reached at the different redshifts by the three planned surveys are clearly shown in Figure~\ref{lir_z_smi}.

According to the model considered here (reproducing the latest {\em Herschel} results), as we go deeper in flux, the higher-$z$ tail becomes increasingly dominated by star-forming galaxies hosting an obscured AGN ({\tt SF-AGN(SB)}; orange points and dot-dashed line in Figures~\ref{lir_z_smi} and \ref{zdistr_smi}, respectively). These galaxies will eventually outnumber the AGN-dominated objects ({\tt AGN1} and {\tt AGN2}; blue and magenta), which prevail at the brighter fluxes at high-$z$.
Therefore, the UDS at high-$z$ will be dominated by elusive/obscured AGN hosted
by SF galaxies (likely to be the obscured phase preceding the AGN-bright epoch), making it even more important to develop a reliable tool for selecting and identifying this kind of 
object and for separating the SF from the AGN contribution. 
This scenario is based on models, while it is necessary to measure the high-$z$ IR populations through photometric surveys. In order to investigate the nature of the faint/high-$z$ mid-IR sources, 
we will need to follow up the SMI/CAM 34-$\mu$m detections in other bands such as optical, far-IR, sub-mm and X-ray. 

\begin{figure}
\begin{center}
\includegraphics[width=\columnwidth]{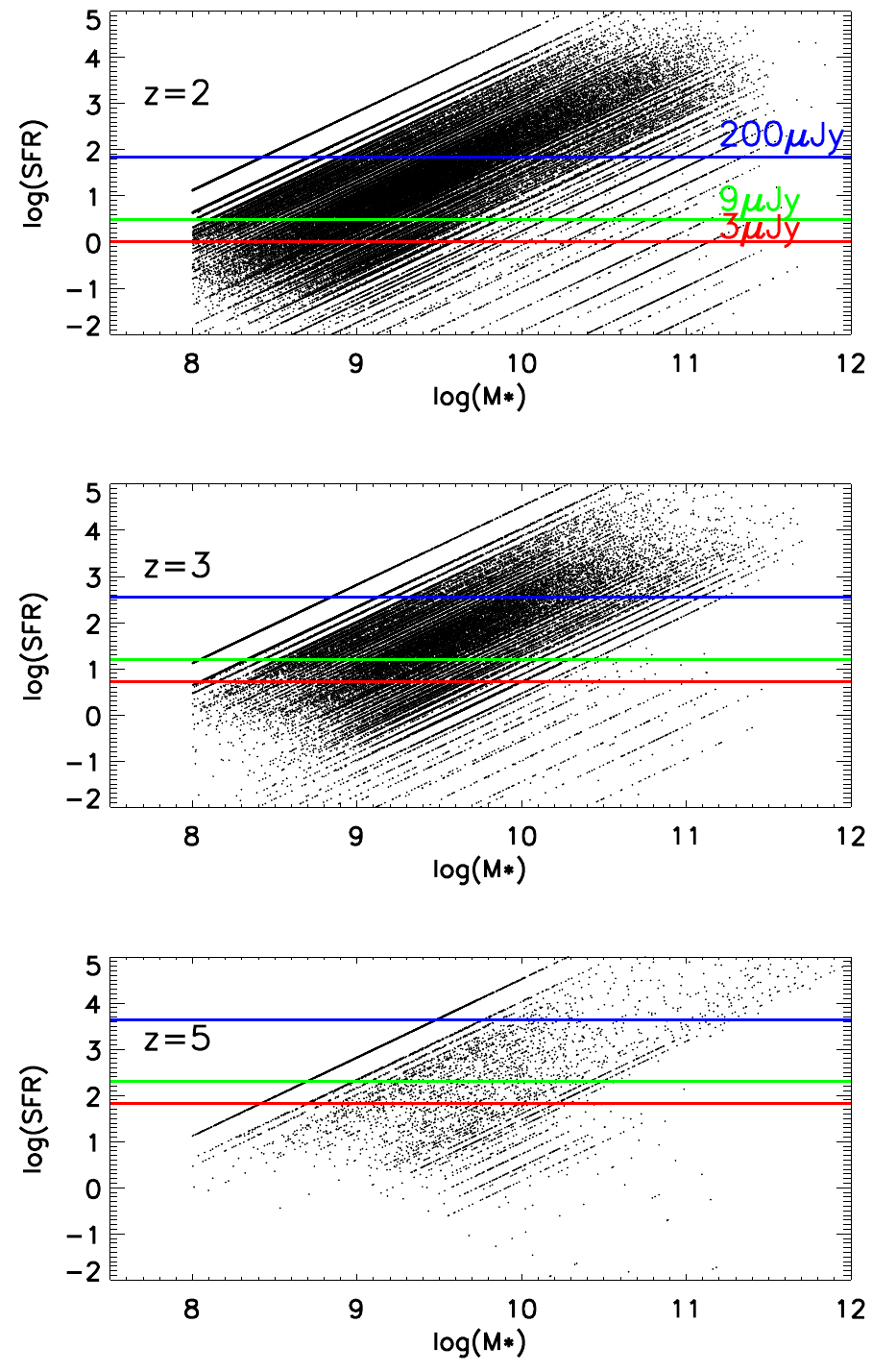}
\caption{The SFR versus stellar mass relation (MS) at $z$$=$2, 3 and 5, as drawn
from the COSMOS survey (Laigle et al. 2016).
Small black dots have been selected from the COSMOS sample in bins of
photometric redshift, with SFR computed from a standard optical/UV
SED-fitting procedure. We show with horizontal solid lines the
predictions for the level of SFR reachable by the
proposed SPICA photometric surveys at $34\,\mu$m. Different colours mark
the three depths for the deep (red, $3\,\mu$Jy)
medium (green, 9 $\mu$Jy) and shallow (red, $200\,\mu$Jy) surveys.
Clearly, only with the deeper integrations will SPICA be able to probe
the bulk of the MS SF galaxies
(at least up to $z$$=$3). At larger redshifts ($z$$>$3), we will rely on
statistical stacking techniques.}
\label{MS_spica}
\end{center}
\end{figure}

In Figure~\ref{MS_spica} we show the SFR-stellar mass relation (also called the ``main-sequence'', MS), found by several authors for 
star-forming galaxies over a wide range of redshifts and luminosities (e.g., \citealt{noeske07, elbaz07, daddi07, rodighiero11}),
as derived for the COSMOS survey (\citealt{laigle16}) in bins of photometric redshift up to $z$$=$5. The SFRs have been computed using a 
standard SED-fitting method in the UV/optical domain (by assuming discrete values of the SF histories, extinction and stellar ages,
responsible for the parallel stripes).
The horizontal lines show the accessible SFRs reaching 3, 9 and $200\,\mu$Jy with a 34-$\mu$m survey
(limiting fluxes of the UDS, DS and SS, respectively). With the UDS (red lines) we will be able to study the SF population lying along the MS 
at $z$$=$3, and most of them at $z$$=$5. The DS will be able to recover most of the SFR--M* relation for SF galaxies at $z$$=$3, while the SS 
already at $z$$\lsimeq$2 will be useful for studying the populations lying above the line, corresponding to objects with more intense SFRs.

The proposed SPICA photometric survey, as planned above, would be complementary to those expected to be carried out with JWST, ALMA and the Square Kilometre Array (SKA, operating at
radio wavelengths). 
While the SMI/CAM will lack the very high resolution 
of JWST-MIRI and ALMA, it will be much more efficient in mapping large fields, detecting large numbers of mid-IR sources.
ALMA is ideal for carrying out high resolution observations of high-$z$ sources (detecting sources easily up to $z$$=$10), given its extremely high sensitivity and the k-correction becoming negative at millimetre (mm) wavelengths. However, given its small FoV (i.e., around 1$^{\rm \prime}$$\times$1$^{\rm \prime}$ at 115 GHz, Band 3), ALMA is not efficient for observing large areas of the sky. It is, however, crucial for constraining the peak of dust re-emission due to SF (and measuring the SFR) in high-$z$ sources.

Deep mid-IR surveys are planned with JWST, which in just a few years time will perform high-sensitivity observations of galaxies at near-/mid-IR wavelengths with high resolution imaging spectroscopy. 
These observations will be already completed by the time SPICA will fly, and could be used to complement and test the validity of the SPICA survey data. However, analogously to ALMA, JWST-MIRI will observe only small areas of the sky (i.e., the FoV of the mid-IR instrument MIRI is only 1.25$^{\prime}\times1.88^{\prime}$). 

The SKA Phase 1 multi-tiered survey envisaged by \citet{prandoni15} at 1--10 GHz is expected to reach SFR levels comparable to those of SPICA and similar redshifts. These observations, already 
in hand at the time SPICA is expected to fly, combined with the SMI/CAM survey data will allow to determine the validity and to study the evolution of the radio-IR correlation up to $z$$\simeq$6. SMI/CAM
will provide identification for the dusty SKA 1 sources and a crucial photometric point probing warm dust, allowing a better separation between AGN and star-forming galaxies.  

Thanks to the large FoV (10$^{\prime}$$\times$12$^{\prime}$) and the mapping speed of over 100 times faster than its mid-IR predecessor JWST (see \citealt{Kaneda17}), 
SMI/CAM on SPICA will easily cover significantly wider areas of the sky. 
The main advantages of a larger area lie in the detection of  statistically significant numbers of sources spanning a wide range of redshifts and luminosities 
(crucial for evolutionary studies, e.g., the LF at different redshifts), the discovery of rare objects and the coverage of large volumes of the sky needed for large-scale structure studies. 

\begin{figure}
\begin{center}
\includegraphics[width=\columnwidth]{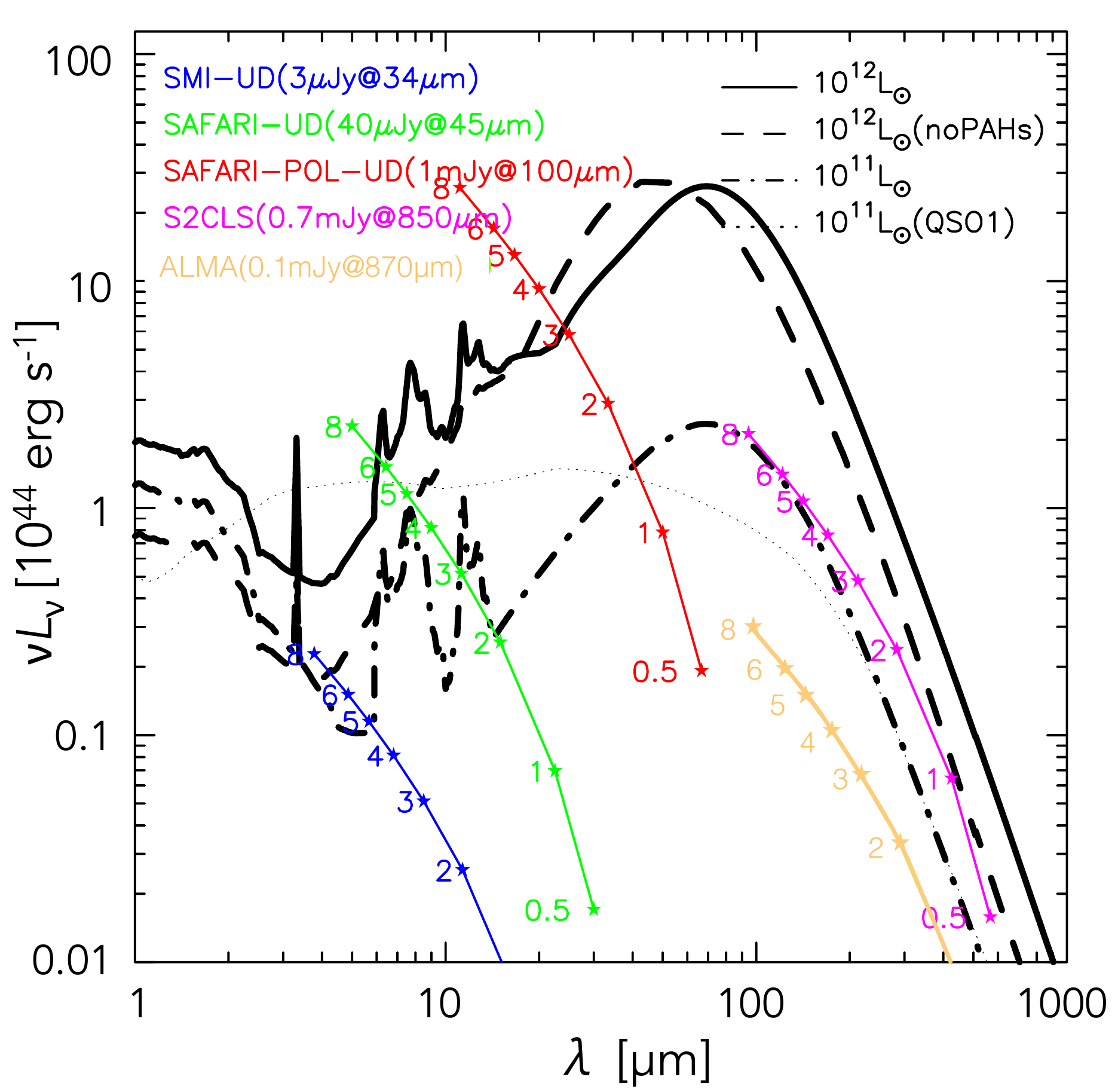}
\caption{Predicted detectability of sources of different SED types (LIRG and ULIRG with and without PAHs, QSO) by SPICA (SMI, SAFARI and SAFARI/POL, compared to ALMA and SCUBA-2.
The coloured curves show the (5$\sigma$ confusion) limits of the different instruments/facilities at different redshifts, with the small stars corresponding to the listed values (blue: SMI; green: SAFARI 
photometry at $45\,\mu$m; red: SAFARI/POL at 100 $\mu$m; magenta: SCUBA-2 at $850\,\mu$m; orange: ALMA Band 8 at $870\,\mu$m).}
\label{SED_spica}
\end{center}
\end{figure}

In Figure~\ref{SED_spica} we show an example of the detectability of sources of different SED types (e.g., LIRG, ULIRG with and without PAHs, and QSO) by SPICA 
(SMI/CAM, SAFARI in photometric mode and SAFARI-POL), ALMA and SCUBA-2. Note that these are all local template SEDs, thus they might be different at high-$z$; however, the indication from the figure is that SMI/CAM will be able to detect sources with these SEDs up to high redshifts ($z$$>$6) in the UDS, while SAFARI-POL will be mostly useful in the
SS, since it will reach $z$$\simeq$3--4 only for the ULIRGs. 
The coloured curves in Figure~\ref{SED_spica} show the (confusion) limits of the different instruments at different redshifts (as described in the figure caption). 
It is evident how such SEDs will be easily detected by ALMA up to very high redshifts (i.e. $>$8); this is a crucial point
for those sources not identified by any other means (e.g., through SAFARI spectroscopy), which will be fully characterised by follow-up observations with ALMA 
and the next generation of ELTs.


\section{BROADBAND FOOTPRINTS OF DISTANT, OBSCURED AND COMPTON-THICK AGN}
\label{AGN}
The synergy and complementarity of the capabilities of both SPICA and ATHENA, which are foreseen to operate during the same period (launch at the end of the 2020s) would 
probe the full cosmic history of BH accretion, regardless of obscuration.
The new window in the 30--$37\,\mu$m range, that can only be explored by SPICA, will be crucial for detecting the mid-IR excess of CT sources up to $z$$\simeq$6--7 
(sampling warm dust at rest-frame 4--$5\,\mu$m at those redshifts). These AGN are believed to be responsible for most of the power produced by accretion in the Universe 
(e.g., Gilli et al. 2007; Treister et al. 2009) and therefore are likely to represent a crucial phase in the joint evolution of galaxies and AGN. 
Due to the powerful capability of the 30--$37\,\mu$m band for detecting AGN at high-$z$, a degree-scale SMI photometric survey at $34\,\mu$m is an optimal way to 
answer the following key questions: What were the formation sites and hosts of the first SMBHs at the highest accessible redshifts ($z$ up to 6--7)? 
Did SMBHs form from small, low-mass seeds ($\sim$10$^{2}$M$_{\odot}$) and accrete above the Eddington limit, or did they form from intermediate-mass seeds ($\sim$10$^{2-5}$M$_{\odot}$) 
and accrete at or below the Eddington limit? Did they form instead by direct collapse of BHs?

\begin{figure}
\begin{center}
\includegraphics[width=\columnwidth]{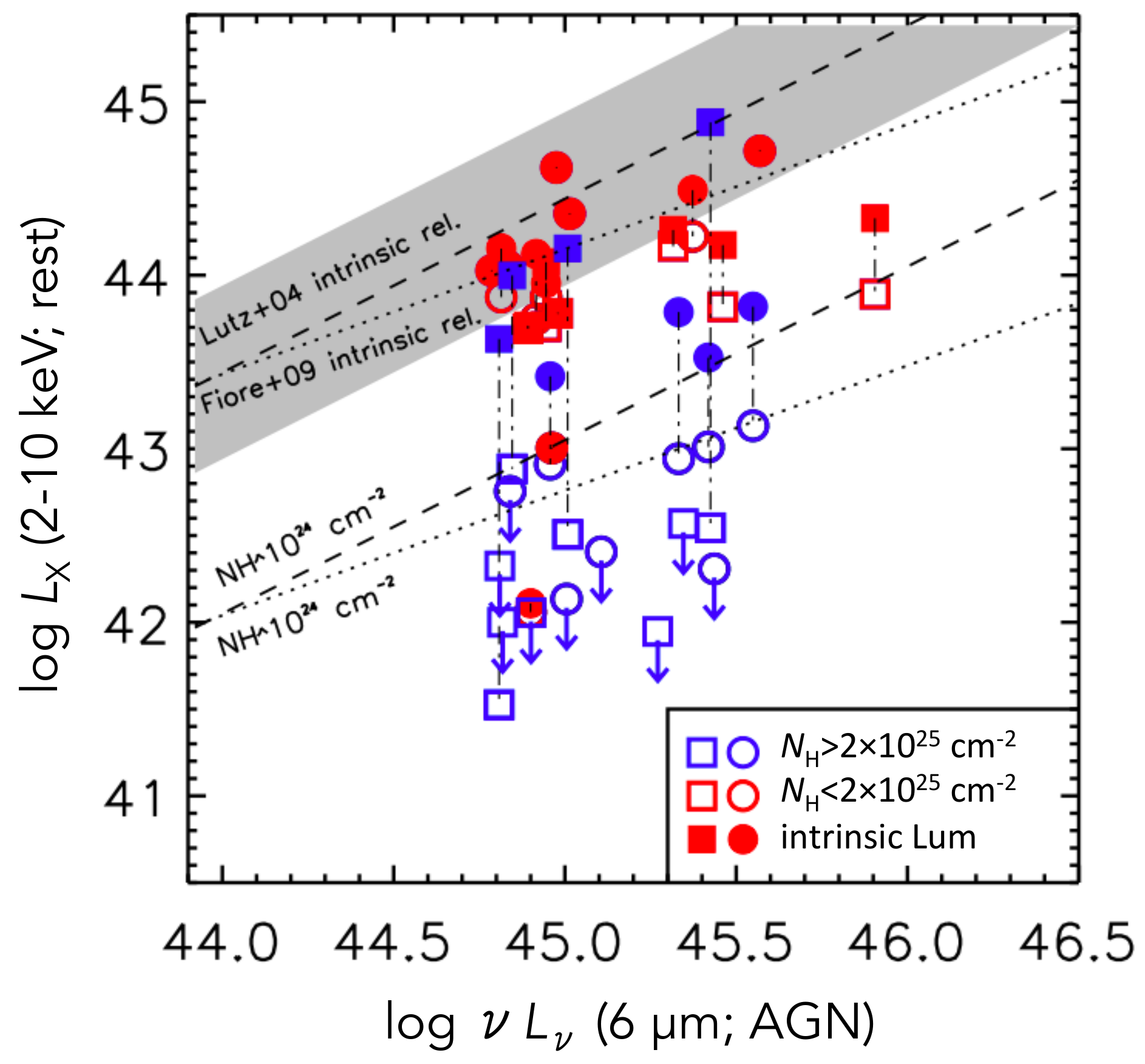}
\caption{X-ray luminosity (2--10 keV rest-frame) versus AGN rest-frame $6\,\mu$m luminosity (in units of erg s$^{-1}$) calculated from the SEDs; the observed X-ray
luminosity (not corrected for absorption) is plotted with open symbols, while the intrinsic luminosity, i.e., corrected for the N$_{\rm H}$ measured from the X-ray
spectra, is plotted with filled symbols. Unobscured and moderately obscured quasars (N$_{\rm H}$$<$2$\times$10$^{23}$~cm$^{-2}$) are plotted in blue, while
the heavily-obscured sources (N$_{\rm H}$$>$2$\times$10$^{23}$~cm$^{-2}$) are plotted in red. Circles and squares correspond to sources in GOODS-N and -S, respectively.
The shaded region represents the scatter of the intrinsic $L_{\rm 6\mu m}$--$L_{\rm X}$ relation found by \citet{lutz04}, shown as dashed line, while the dotted line represents the relation found by \citet{fiore09}.}
\label{XrayIR}
\end{center}
\end{figure}
A significant fraction (about 30\%) of the mid-IR selected AGN at $z$$\simeq$2 are undetected in the current deep X-ray surveys 
(e.g., \citealt{fiore08,delmoro16}) and not even the future X-ray missions such as ATHENA (i.e., the ESA L2 mission expected to fly in 2028; \citealt{nandra13})
will be able to reveal all of them (e.g., still $\sim$20\% of all the SMBHs and up to 50\% of the CT AGN could be X-ray silent and identifiable only through mid-IR observations; \citealt{comastri15}). 
Figure~\ref{XrayIR}, from \citet{delmoro16}, shows the X-ray luminosities against the 6-$\mu$m luminosities 
(considered as a proxy of the AGN luminosity) for a sample of mid-IR luminous quasars at 1$<$$z$$<$3, identified through detailed SED analyses in 
the GOODS-{\em Herschel} fields; despite being very bright in the mid-IR band, about 30\% of these quasars
are not detected in the extremely deep {\em Chandra} X-ray data available in these fields (even the reflection 
spectrum and the usually strong Fe K$_{\alpha}$ line at 6.4 keV are buried) and a large fraction are found to be CT (24--48\%).
This would mean that a large part of the cosmic X-ray background arises from a population mostly 
undetectable by current and future X-ray surveys, with nuclear obscuring column densities even higher than N$_{\rm H}$ $\simeq$10$^{25}$cm$^{-2}$ \citep[e.g.,][]{comastri15}. 
Current estimates of the fraction of these CT systems range from a few percent to over 50\% of the overall obscured AGN population \citep[e.g.,][]{brightman11b}. 

\begin{figure}
\begin{center}
\includegraphics[width=\columnwidth]{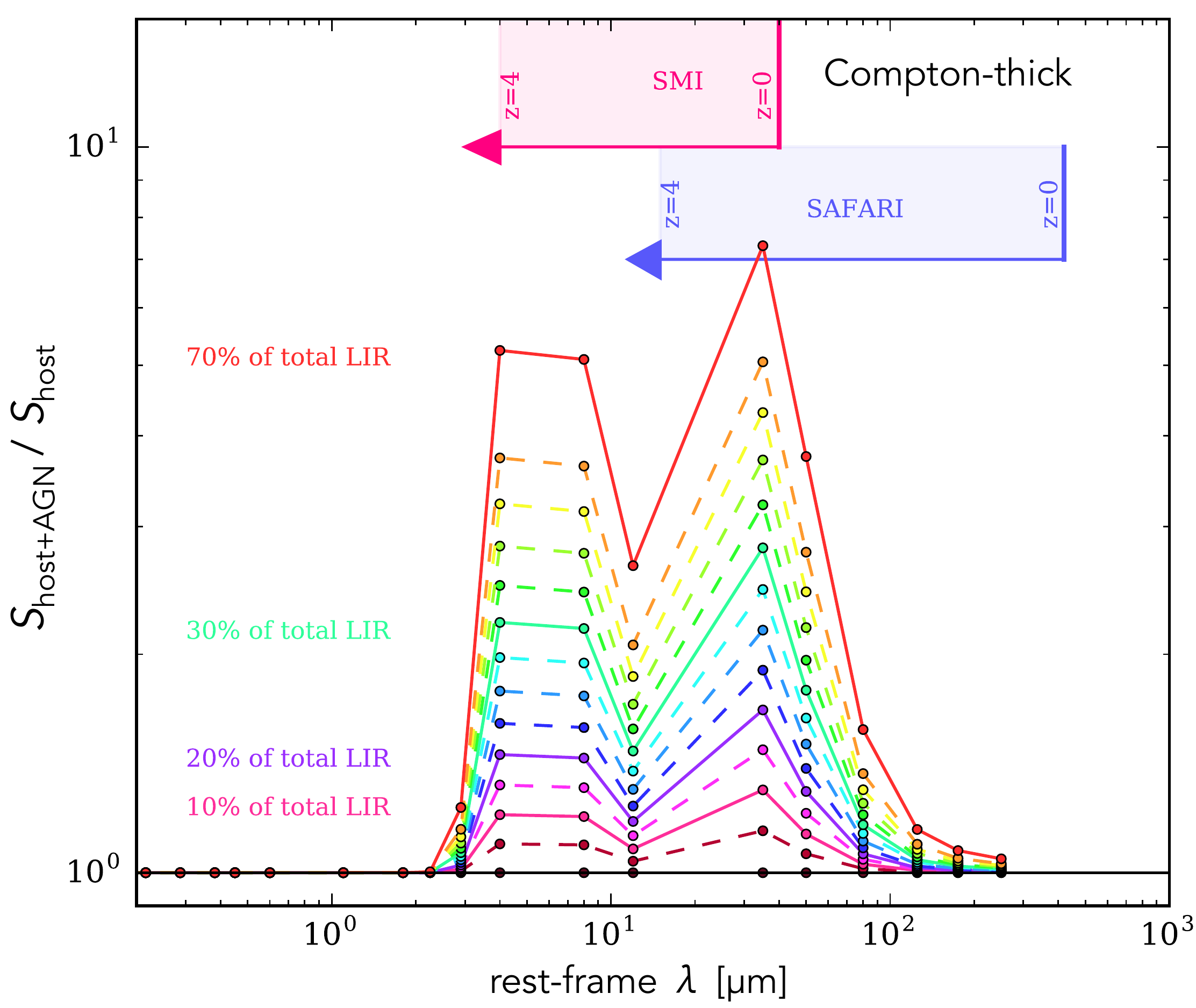}
\caption{Ratios between the SEDs of a galaxy hosting a CT AGN and the SED of the same galaxy without AGN emission.
The different coloured lines indicate different contributions of the AGN luminosity to the total IR luminosity, from 10\% to 70\%.
The spectral regions probed by SMI/CAM (30--$37\,\mu$m) and SAFARI/POL (100--$350\,\mu$m), in photometric mode, from redshift 0 to 4, 
are indicated with the red and blue regions respectively, 
and correspond to the rest-frame spectral range where we expect the AGN to impact the most its host galaxy SED. 
SPICA (SMI $+$ SAFARI) will cover the entire mid-/far-IR bolometric output of these sources, disentangling the AGN 
from the SF galaxies (what {\em Herschel} could not do without a mid-IR channel).}
\label{fluxratio}
\end{center}
\end{figure}

Deep mid-IR (spectro)-photometric surveys will push the search for heavily obscured AGN to higher redshifts and significantly lower sensitivity levels than ever achieved before, which, 
together with multi-wavelength ancillary information and 
IR spectroscopy, will provide complete samples of heavily-obscured and CT AGN. 
In fact, most of the CT sources missed by the deepest X-ray surveys reveal themselves in the mid-IR, thanks to the warm emission from the 
circum-nuclear dust (i.e., torus) around the central 
accretion disk. This warm emission manifests itself as a mid-IR excess, with the emission from the AGN-heated dust significantly adding to the mid-IR SED of a galaxy. 
Indeed, in Figure~\ref{fluxratio} we show the ratio between a set of simulated SEDs in which we added the contribution of a CT AGN to the SED of the host galaxy \citep[see][for more details]{ciesla15}.
The candidate CT AGN could be selected from deep 34-$\mu$m photometric observations, then followed-up by SAFARI, with their fine-structure lines (if detectable in high-resolution spectroscopy) 
allowing a precise measurement of their redshifts, an accurate determination of the origin of the energy powering the IR emission (SF versus AGN) and, finally, a reliable evaluation of the bolometric 
luminosity of the AGN emission. 
The contribution of the AGN to the total IR luminosity varies from 10 to 70\%.
It is clear that the spectral regions in which we we expect the CT AGN to affect the most of the host galaxy SED is between 3--12\,$\mu$m and 20--100\,$\mu$m in the rest frame. 
As shown on Figure~\ref{fluxratio}, these ranges are perfectly probed by SMI and SAFARI.
By selecting mid-IR sources with faint near-IR and optical emission, \citet{daddi07} and \citet{fiore08} suggest that several mid-IR excess sources in the Chandra Deep Field-South can be obscured AGN at 
1$\leq$$z$$\leq$3, on the basis of stacking techniques applied to X-ray wavelengths. 
The combination of both the 34-$\mu$m (SMI) and the 100-$\mu$m (SAFARI) data points can be used with models of circum-nuclear dust emission (e.g., \citealt{efstathiou95, fritz06, nenkova08, honig10, feltre12}) 
and star-forming galaxy templates to separate the part of the luminosity originating from accretion onto a black hole from that due to SF. This is an advantage over the information
provided by {\em Herschel} and the JWST, which will miss the mid- and far-IR channels respectively.
Performing this kind of survey for the crucial $z$$\sim$1--4 epoch over degree-size fields, with a single mission in a reasonable amount of time, is beyond the reach of any current or planned facilities
other than SPICA. 

The AGN population currently known at $z$$\simeq$6--7 consists of luminous optical quasars (e.g., \citealt{fan03}, \citealt{gallerani17}) and although these objects host the most massive BHs 
($>$10$^9$~M$_{\odot}$) in the Universe, they are extremely rare. 
At these redshifts, typical AGN, which are of lower luminosity and often obscured and/or diluted by their host galaxy emission, remain largely undiscovered. 
X-ray selection has provided the most robust AGN samples to date, but finding the most obscured objects has proved challenging. 
To date, no obscured AGN have been identified at $z$$>$5 (the most distant known is at $z$$=$4.75; \citealt{gilli14}).
Recently some AGN have been identified in low metallicity systems, lacking X-ray emission (\citealt{simmonds16}). 
Therefore, a complete census of the AGN population and a complete BHARD estimate, would require a selection complementary to X-ray (e.g., in the IR).
SPICA will be able to detect heavily CT AGN, i.e., those that are fully or almost fully enshrouded and embedded in dust, so that even the reflection 
spectrum and the Fe K$_{\alpha}$ line at 6.4 keV lie buried and undetected at X-ray energies. 

The relative abundance of weak and powerful AGN (independently of obscuration) as a function of cosmic time and host galaxy properties, is a vital ingredient of any attempt to 
clarify the entangled history of the galaxies and their active nuclei.
At low AGN luminosities, where the host galaxy emission may limit the mid-IR target selection, there are obvious synergies with ATHENA, which is expected to
reveal hundreds of mildly CT AGN: IR spectroscopy with SPICA would allow us to fully characterise these sources. Among the scientific goals of  ATHENA
is, in fact, the detection of a few hundreds of $z$$>$6 AGN -- which could be followed-up by deep SPICA SMI photometry and SAFARI spectroscopy -- and a few tens of CT AGN 
at 1$<$$z$$<$4 (\citealt{aird13}). The CT AGN are expected to be hosted in galaxies as faint as A mission
$\simeq28\,$mag (AB) in the optical/near-IR. At these depths the source density is expected to be about 1 arcsec$^{-2}$ (\citealt{guo13}), corresponding to 3 optical 
sources within the ATHENA WFI FoV/error circle. 
It will therefore be challenging to identify the correct counterpart of the X-ray emitter (although Bayesian methods might help).

\begin{figure}
\begin{center}
\includegraphics[width=\columnwidth]{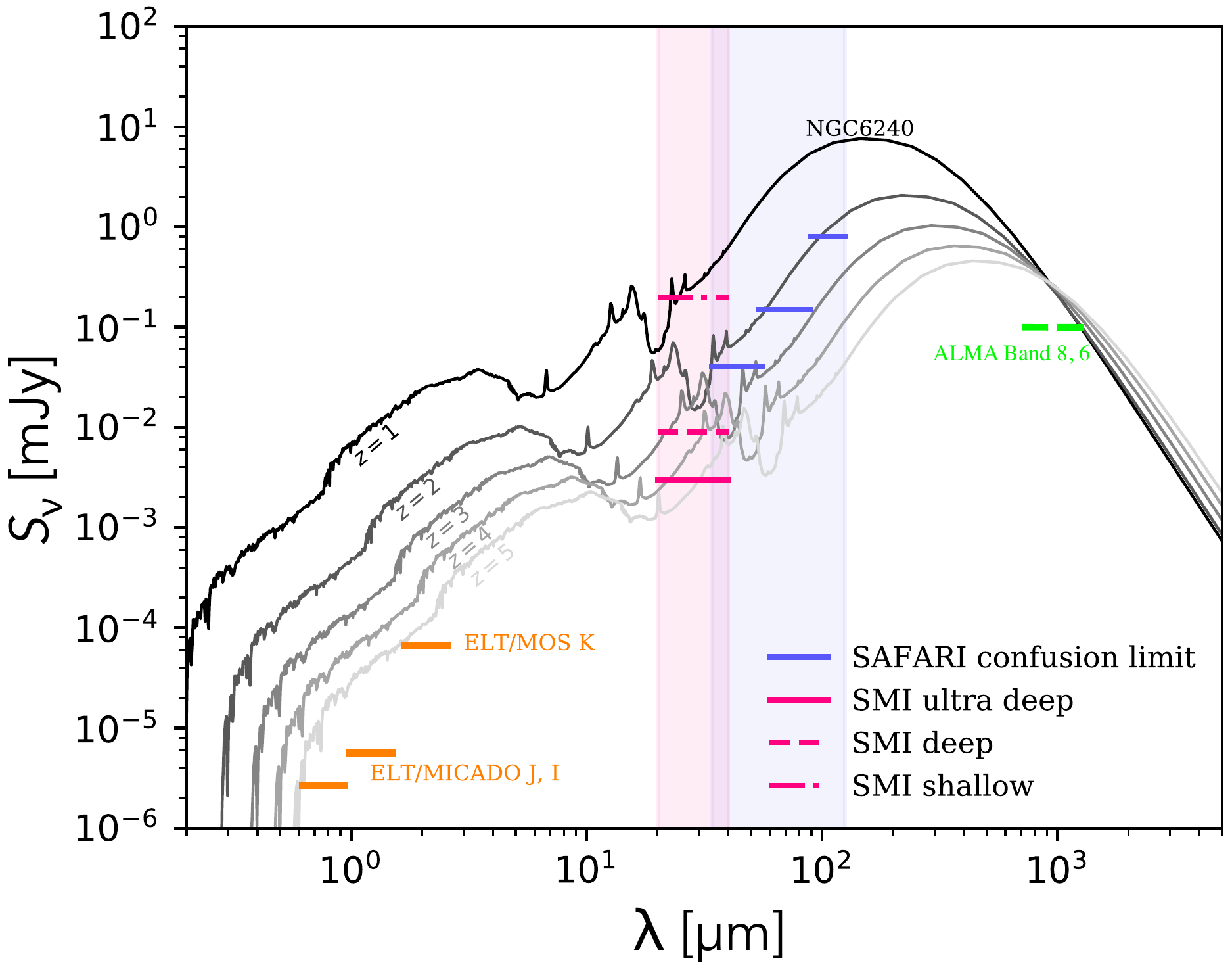}
\caption{Template SED of the local CT AGN NGC6240 (obtained by fitting the observed data with CIGALE; {\tt http://cigale.lam.fr/}),
scaled by redshift up to $z$$=$5. NGC6240 is a LIRG, with $L_{\rm IR}$$=$7$\times$10$^{11}$~L$_\odot$. The pink and blue regions show the wavelength ranges sampled by SMI (30--$37\,\mu$m) and SAFARI (at 45, 72 and $100\,\mu$m, over the bands 34--56 and 54--$89\,\mu$m in photometric mapping, and 75--$125\,\mu$m with SAFARI-POL), respectively, while the
pink horizontal lines represent the limits of the Ultra-Deep, Deep and Shallow reference surveys planned with SMI and described in Section~\ref{photo_survey}. 
The blue horizontal lines show the confusion flux density for a 2.5-m telescope in the SAFARI bands. For comparison, ALMA (3$\sigma$, 5\,hr in Band 8 and 20\,min in Band 6, green horizontal lines), 
ELT/MOS and ELT/MICADO (3\,hr, orange horizontal lines) detection limits are shown.}
\label{agn_sensitivity_spica}
\end{center}
\end{figure}
A survey centred at $34\,\mu$m down to (or below) the estimated confusion limit (like the one described in the previous section) would be sensitive to moderate luminosity 
($L_{\rm IR}$$\sim$10$^{11}$~L$_{\odot}$) obscured AGN out to $z$$\simeq$5--6, and particularly in the 3$\leq$$z$$\lsimeq$6 range, 
where the co-evolution of SF and accretion activity is expected to already be in place. 
It will also be sensitive to lower luminosity/Seyfert-like AGN ($L_{IR}$$\sim$10$^{10}$~L$_{\odot}$) to $z$$\simeq$2 (see Figure~\ref{lir_z_smi}). 
In Figure~\ref{agn_sensitivity_spica} we show an example of the expected SPICA detectability of the local CT AGN NGC6240, classified as LIRG ($L_{\rm IR}$$=$7$\times$10$^{11}$~L$_\odot$): SMI will be able to detect it up to $z$$\simeq$5 (in the UDS), while follow-up observations with SAFARI for such an object will be possible only up 
to $z$$\simeq$2--2.5, even in the SAFARI's less confused photometric bands (i.e., 45 and $72\,\mu$m). ELTs and ALMA will be able to easily detect sources similar to NGC6240 out to $z$$\simeq$5 in a few minutes up to hours, and to higher redshifts by integrating for longer than considered in the plot.

\section{STAR-FORMING GALAXIES AT $z$$\gsimeq$4}
\label{SB}
\begin{figure}
\begin{center}
\includegraphics[width=\columnwidth]{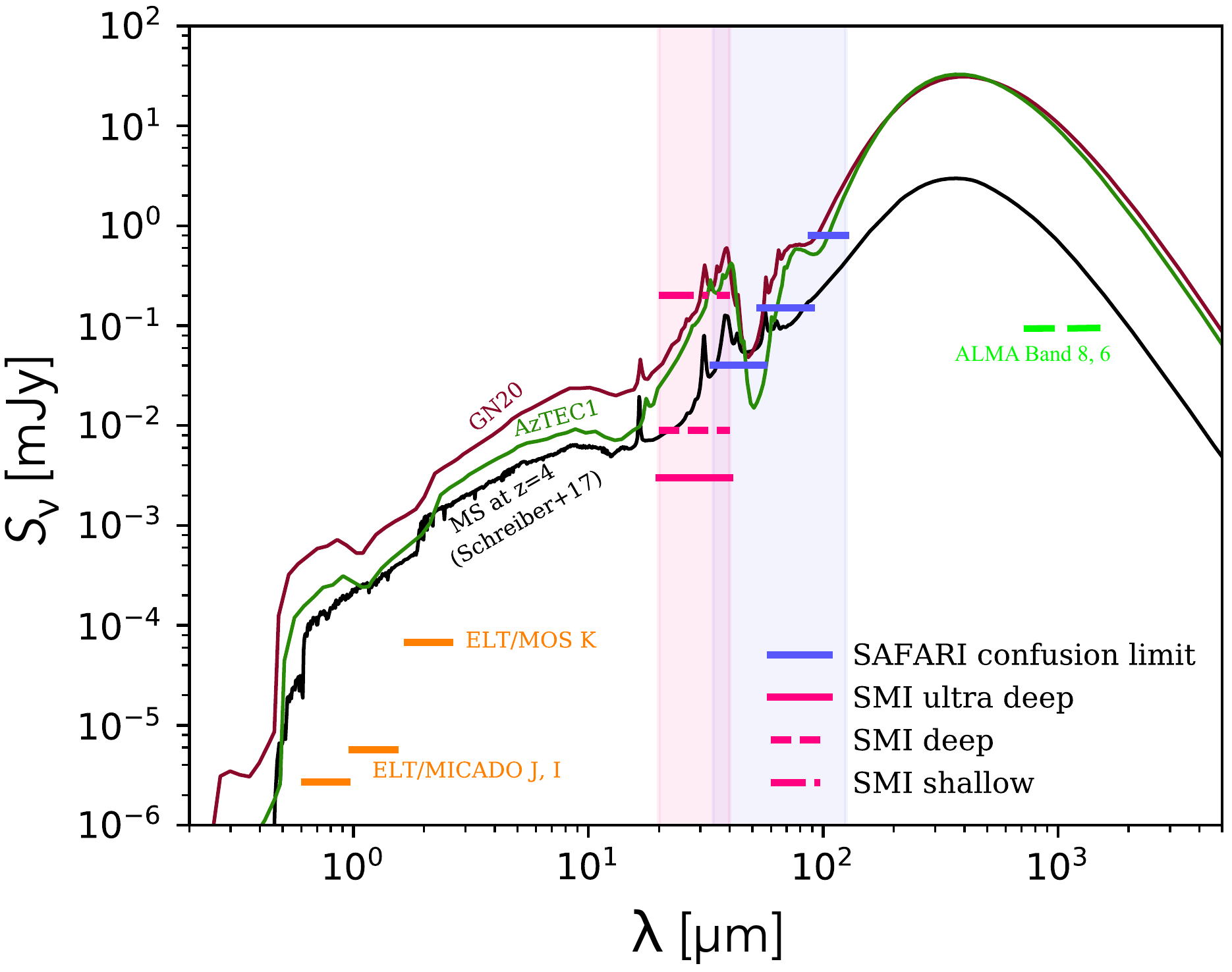}
\caption{Fits to the observed SEDs of: the $z$$\simeq$4 starburst galaxy GN20 (purple); and the $z$$=$4.34 
sub-mm galaxy AzTEC-1 (green). These are both fitted with the models of  \citet{efstathiou09}, and the average $z$$\simeq$4 MS galaxy observed by \citet{schreiber17} with ALMA (black). 
Pink horizontal lines represent the limits of the Ultra-Deep, Deep and Shallow reference surveys planned with SMI and described in Section~\ref{photo_survey}. 
The blue horizontal lines show the confusion flux for a 2.5-m telescope in the SAFARI bands. For comparison, ALMA (3$\sigma$, 5~hr in Band 8 and 20~min in Band~6: 
green horizontal lines), ELT/MOS and ELT/MICADO (3~hr, orange horizontal lines) detection limits are shown.}
\label{GN20}
\end{center}
\end{figure}

Observing distant galaxies in the far-IR/sub-mm is challenging and only the brightest objects, experiencing a burst of SF activity can be directly detected (e.g., \citealt{pope06,capak11,riechers13}).
However, they have extreme SF properties and IR luminosities, but low spatial densities, and are not representative of the high-redshift galaxy population at $z$$=$3--4.
This is, for instance, the case for GN20 (\citealt{daddi09}) at  $z$$=$4.05, one of the best studied sub-mm galaxies at $z$$>$4, showing a 6.2-$\mu$m PAH feature (\citealt{riechers14}), and COSMOS AzTEC-1 at $z$$=$4.3, studied by \citet{yun15}.

To probe the typical properties of the bulk of the galaxy population at $z$$>$3, stacking analyses have been developed to study the average SFR and the gas mass of galaxies not individually 
detected with the deepest \textit{Herschel} images (e.g., \citealt{magdis12,heinis14,schreiber15,bethermin15,tomczak16}).
ALMA has allowed us to detect less extreme galaxies at $z$$\geq$4 (\citealt{capak15,maiolino15,scoville16}), but these samples are mostly based on UV-selected samples, biased significantly against massive galaxies (\citealt{spitler14,wang16}), and lack of statistics. 
Recently, \citet{schreiber17} observed a sample of a hundred massive galaxies ($M^*$$>$5$\times$10$^{10}$~M$_{\odot}$) at $z$$\simeq$4 with ALMA. A third of the sample was detected, revealing that the MS relation of 
galaxies was already in place at this redshift, and they derived an average dust temperature of 40\,K for $z$$=$4 galaxies.
This dust temperature is higher than what was determined for lower redshift galaxies and indeed GN20, i.e. 33\,K (\citealt{magdis12}), implying IR emission peaking at slightly shorter wavelengths. This would make the bulk of high redshift galaxies easily detectable by SPICA.

Figure~\ref{GN20} shows the SED of the typical MS galaxies at $z$$=$4 detected by \citet{schreiber17} at 870\,$\mu$m with ALMA (discussed above), as well as the 
SEDs of the two starbursts GN20 and AzTEC-1. Note that it is likely that some of the MS galaxies that have been considered to derive the average template might contain an AGN, thus this
source cannot be considered as a ``normal'' SF galaxy (as it is referred to).
A survey with SMI/CAM will easily allow us to build a statistically significant sample of $z$$\simeq$3--5 MS galaxies, removing the biases suffered from previous studies of individual extremely luminous objects.
Figure~\ref{GN20} shows that in the SMI/CAM DS (reaching confusion at 34\,$\mu$m), we would be able to detect galaxies with SEDs similar to these templates at least out to $z$$\simeq$4
and beyond (in the UDS). Dusty galaxies like GN20 and AzTEC-1 could be followed up in at least two bands (i.e., 45 and 72\,$\mu$m) by SAFARI in photometric mode at least out to $z$$\sim$4, while a galaxy similar to the average MS template could be detected by SAFARI at 45\,$\mu$m (but not beyond $z$$\sim$4). For the follow up of normal MS star-forming galaxies, ALMA and ELTs will be necessary.
As shown in the plot, ALMA and E-ELT in a very short time (from few minutes to few hours) will reach fluxes significantly fainter than those needed to detect these $z$$=$4 sources, being able to easily identify galaxies with such SEDs up to much higher redshifts and/or fainter luminosities. 

\section{CONCLUSIONS}
A multi-layer photometric survey at 34\,$\mu$m with SPICA-SMI/CAM would enable the observations of galaxies and AGN to unprecedented depth, in an unexplored band 
between the JWST and the {\em Herschel} wavelength ranges. Such surveys will be sensitive enough to detect hundreds of $z$$\sim$6 sources in a relatively small
amount of time ($\sim$240~hours in total) and have an FOV large enough to reveal tens of thousands of galaxies and AGN at $z$$=$3--5. 
Due to the wavelength coverage and the efficiency in observing large areas of the sky, SPICA-SMI/CAM will be unique in its capability to scan the sky at infrared
wavelengths, studying the evolution of dusty galaxies with unprecedented statistics and depth and discovering new classes of objects, 
either faint, or rare or peculiar, i.e., sources not previously detected in any other bands. 
A UDS of 0.2~deg$^{2}$, 3~times below the estimated confusion, a DS of 1~deg$^{2}$ to confusion and an SS of 600~deg$^{2}$ to 0.2~mJy can be 
considered as the best strategy for studying galaxy evolution with SPICA photometry, 
covering the faint-end, the knee, and the bright end of the LF up to high redshifts ($z$$\simeq$6) with statistically significant samples of sources (at any redshifts and luminosities).
The proposed ``wedding-cake'' survey should be performed on popular and extensively observed fields, to take advantage of the great wealth of multi-wavelength information 
already available. Specific follow-up studies with ELTs and ALMA will be performed for sources lacking any identification in existing deep images and catalogues.
The SPICA photometric surveys will be the only study able to derive the obscured SFRD and BHARD with large statistical samples up to the reionisation epoch, 
an epoch for which the only information available to date is derived using rest-frame UV observations, which are subject to large systematic uncertainties due 
to the presence of dust. These surveys will be the only way to obtain a complete
census of highly obscured (e.g., CT) AGN and to shed light on the amount of dust in galaxies at $z$$>$3--4, as well as to derive important clues about the evolution of galaxies and dust.

\begin{acknowledgements}
This paper is dedicated to the memory of Bruce Swinyard, who initiated the SPICA project in Europe, but sadly passed away on 22 May 2015 at the age of 52. 
He was ISO-LWS calibration scientist, {\em Herschel}-SPIRE instrument scientist, former European PI of SPICA and design lead of SAFARI.
This mission would not have been possible without the SAFARI consortium and the entire SPICA team.

We thank the anonymous referee, whose useful comments and suggestions helped improving this paper.
\end{acknowledgements}

\bibliographystyle{pasa-mnras}

\bibliography{mybibliography}

\end{document}